\documentclass[11pt]{article}

\usepackage{amsmath}
\usepackage{amssymb}
\usepackage{amsthm}
\usepackage{amsfonts}
\usepackage{graphicx}
\usepackage{textcomp}
\usepackage{color}
\usepackage[backref=page]{hyperref}
\usepackage{stmaryrd}
\usepackage{tikz-cd}
\usepackage{setspace}
\usepackage[utf8]{inputenc}
\usepackage[english]{babel}
\usepackage{mathrsfs}
\usepackage{quiver}
\usepackage[a4paper, total={5.7in, 9.5in}]{geometry}
\usepackage{cite}
\usepackage{comment}

\let\OLDitemize\itemize
\renewcommand\itemize{\OLDitemize\addtolength{\itemsep}{-8pt}}

\numberwithin{equation}{section}

\def\sideremark#1{\ifvmode\leavevmode\fi\vadjust{\vbox to0pt{\vss% the remark
			\hbox to 0pt{\hskip\hsize\hskip1em%                          will appear only
				\vbox{\hsize3cm\tiny\raggedright\pretolerance10000%          on the side
					\noindent #1\hfill}\hss}\vbox to8pt{\vfil}\vss}}}%
\begin{document}

\include{preamble}

\begin{titlepage}

\title{Drinfel'd Doubles, Twists and All That... in Stringy Geometry \\ and M Theory}

\author{Aybike Çatal-Özer$^a$, Keremcan Doğan$^{a, b}$, Cem Yetişmişoğlu$^{a, c \dagger}$\footnote{E-mails: ozerayb[at]itu.edu.tr, keremcandogan[at]gtu.edu.tr, yetismisoglu[at]itu.edu.tr} \\ 
\small{$^a$ Department of Mathematics, İstanbul Technical University, Maslak, İstanbul, Turkey} \\
\small{$^b$ Department of Mathematics, Gebze Technical University, Cumhuriyet, Kocaeli, Turkey} \\
\small{$^c$ School of Engineering and Natural Sciences, İstanbul Medipol University, Kavacık, Beykoz, Turkey} \\
\small{$^{\dagger}$Corresponding author}}

\date{}

\maketitle

\begin{abstract}

\noindent Drinfel'd double of Lie bialgebroids plays an important role in T-duality of string theories. In the presence of $H$ and $R$ fluxes, Lie bialgebroids should be extended to proto Lie bialgebroids. For both cases, the pair is given by two dual vector bundles, and the Drinfel'd double yields a Courant algebroid. However for U-duality, more complicated direct sum decompositions that are not described by dual vector bundles appear. In a previous work, we extended the notion of a Lie bialgebroid for vector bundles that are not necessarily dual. We achieved this by introducing a framework of calculus on algebroids and examining compatibility conditions for various algebroid properties in this framework. Here our aim is two-fold: extending our work on bialgebroids to include both $H$- and $R$-twists, and generalizing proto Lie bialgebroids to pairs of arbitrary vector bundles. To this end, we analyze various algebroid axioms and derive twisted compatibility conditions in the presence of twists. We introduce the notion of proto bialgebroids and their Drinfel'd doubles, where the former generalizes both bialgebroids and proto Lie bialgebroids. We also examine the most general form of vector bundle automorphisms of the double, related to twist matrices, that generate a new bracket from a given one. We analyze various examples from both physics and mathematics literatures in our framework.
\end{abstract}

\vskip 2cm


%\tableofcontents

\thispagestyle{empty}

\end{titlepage}

\maketitle

%%%%%%%%%%%%%%%%%%%%%%%%%%%%%%%%%%%%%%

\section{Introduction and Motivation}
\label{s1}

String and M theories both exhibit  features absent in the theories of particle fields.  Extended worldvolumes of fundamental objects lead to different notions of stringy duality symmetries. The presence of dualities suggests a new version of geometry underlying these theories; a stringy geometry. In this new geometric setup, it is possible to combine spacetime symmetries and higher-form gauge transformations. For example,  the gauge transformations of the Kalb-Ramond field and diffeomorphisms can be packaged into a generalized gauge transformation of the generalized metric \cite{hull2009gauge} in Double Field Theory (DFT), which aims to provide a T-duality covariant formulation \cite{hull2009double, Hohm:2010pp, zwiebach2012doubled, Hohm:2011dv, Hohm:2013vpa}  by doubling the spacetime coordinates. The extra coordinates of DFT are associated to the winding modes of the string which can wrap up the compact dimensions. The simplest case of T-duality is then given by exchanging the winding and momentum modes, which can be elevated to an $O(d, d)$ transformation of generalized metric on the doubled spacetime \cite{giveon1994target}. When a certain constraint ensuring the halving of the number of coordinates is imposed, the fields and gauge parameters in DFT are sections of (various tensor products) of the generalized tangent bundle
\begin{equation}
    TM \oplus T^*M \, .
\end{equation}
Again under the same constraint, the generalized gauge transformations of DFT close under the Courant bracket \cite{hull2009gauge}. Equipped with this bracket, the generalized tangent bundle becomes a Courant algebroid \cite{courant1990dirac, liu1997manin}; in fact it is called the standard Courant algebroid. Courant algebroids, and more generally algebroids, provide a useful framework for generalizations of geometric structures including the ones motivated by string and M theories. Courant algebroids  first appeared as the Drinfel'd double of Lie bialgebroids \cite{liu1997manin}, which  are defined as a pair of dual Lie algebroids \cite{pradines1967theorie} satisfying a certain compatibility condition \cite{mackenzie1994lie}. This doubled structure is key to understanding T-duality and its generalizations. For example Poisson Lie T-duality, which includes abelian and non-abelian T-dualities as particular examples, can be understood in terms of  decompositions of the same Drinfel'd double into different Manin triples \cite{klimvcik1995dual, Hassler:2017yza}.

For consistency, string and M theories require extra dimensions, which necessitates a dimensional reduction procedure. In the low energy limit, the $O(d,d)$ T-duality symmetry also appears as a ``hidden symmetry'' in the $11-d$ dimensional supergravity theory arising from the toroidal reduction of 11 dimensional supergravity \cite{samtleben202311d}. The consistency of the lower dimensional effective theory brings restrictions on the nature of extra compact dimensions. For example, requirement of $N = 2$ supersymmetry in 4 dimensions arising from a compactification of Type II string theories forces the internal manifold to be Calabi-Yau, when there are no fluxes. Turning the NS and RR fluxes on, supersymmetry requirement imposes conditions on the internal manifold, which is best described in terms of the structure group of the generalized tangent bundle of the internal manifold \cite{grana2017string, hitchin2010lectures}. 
When T-duality is taken into account, fluxes come up as four types in two categories: geometric $f, H$ and non-geometric $Q, R$ which are crucial for string phenomenology. For instance they can be considered as possible sources for branes, and their presence might be used for moduli stabilization. 
All of these ideas, and in particular the concept of non-geometry, are nicely explained in the review \cite{plauschinn2019non} and references therein. Under (generalized) T-duality, the fluxes fit into the following chain \cite{shelton2005nongeometric}:
\begin{equation}
    H_{a b c} \longleftrightarrow f^{a}{}_{ b c} \longleftrightarrow Q^{a b}{}_{c} \longleftrightarrow R^{a b c} \, .
\end{equation}

More relevant to the purposes of this paper, these fluxes can be regarded as the structure functions/constants of Courant algebroids important in describing stringy geometric structures. For example, it is known that the bracket on any exact Courant algebroid is equivalent to twisted Dorfman bracket on the generalized tangent bundle. On a frame, the structure functions of this bracket are exactly the geometric $f$ and $H$ fluxes, and the axioms to be satisfied by the bracket ensure that the structure functions must obey exactly the same conditions as the Bianchi identities to be satisfied by the geometric string fluxes. Importantly for Poisson Lie T-duality, the double of a Lie bialgebroid has a bracket which realizes $f$ and $Q$ fluxes and again the required Bianchi identities to be satisfied by the $Q$ flux are ensured by the axioms to be satisfied by the bracket. More generally, all of the four fluxes fit into a Kaloper-Myers algebra \cite{kaloper1999dd}:
\begin{align}
    [e_a, e_b] &= f^c{}_{a b} e_c + H_{a b c} e^c \, , \nonumber\\
    [e_a, e^b] &= Q^{b c}{}_{a} e_c + f^b{}_{ a c} e^c \, , \nonumber\\
    [e^a, e^b] &= R^{a b c} e_c + Q^{a b}{}_{c} e^c \, .
\label{km}
\end{align}
These commutation relations give the most general gauge algebra coming from a generalized duality twisted reduction \cite{aldazabal2011effective}, and they can be realized by a Roytenberg bracket of the form
\begin{equation}
    [U + \omega, V + \eta]_E = [U, V]_{TM} + \mathcal{L}^*_{\omega} V - \iota^*_{\eta} d^* U + R(\omega, \eta) \oplus [\omega, \eta]_{T^*M} + \mathcal{L}_U \eta - \iota_V d \omega + H(U, V) \, . \label{roytenbergbracket}
\end{equation}

\noindent One of the main objectives of this paper is to analyze generalizations of this bracket, which we hope will lead to a better understanding of string and M theory fluxes related to U-duality. In the bracket (\ref{roytenbergbracket}), $H$- and $R$-twists are present, so that both $TM$ and $T^*M$ are not subalgebroids of $E = TM \oplus T^*M$ as opposed to the Lie bialgebroid case. Nevertheless, this bracket too furnishes the generalized tangent bundle with a Courant algebroid structure under certain conditions \cite{roytenberg2002quasi}. These conditions define the notion of proto Lie bialgebroids which extends Lie bialgebroids \cite{mackenzie1987lie} into the twistful case. One can associate a topological field theory, called Courant sigma model, to any Courant algebroid \cite{roytenberg2007aksz}. The consistency of this model is governed by the Bianchi identities which are guaranteed by the properties of Courant algebroids \cite{geissbuhler2013exploring, blumenhagen2012bianchi, ikeda2003chern}. Existence of twists in this setting may lead to relaxation of these properties. 
 
This relaxation can be also captured by other relevant but different notions of twists. For instance topological open membranes related to twisted Poisson structures \cite{klimvcík2002wzw, vsevera2001poisson} are studied in \cite{hofman2004bv} where their connections to Lie and Courant algebroids are investigated as well. Another notion of twist (matrix) appears in the generalized vielbein formalism, which plays a fundamental role  for understanding non-abelian T-duality, Yang-Baxter deformations and more generally  Poisson-Lie T-duality in the framework of DFT \cite{ccatal2019non, Catal-Ozer:2019tmm, sakatani2019type, Hassler:2017yza, aldazabal2011effective}. One can also define a notion of twist for a quasi Lie bialgebroid, where the $R$-twist is absent, with respect to a bivector \cite{roytenberg2002quasi}. The Poisson condition for the bivector is equivalent to the condition that the twisted algebroid structure remains a quasi Lie bialgebroid \cite{kosmann2011poisson}. Such twists sustain a powerful technique for generating new algebroid structures. \textit{Quasi, twisted, and all that in Poisson geometry and Lie algebroid theory} are nicely explained in the seminal paper of Yvette Kosmann-Schwarzbach \cite{kosmann2005quasi}.

Extending these ideas for the case of U-duality requires further work to which we hope to contribute with this paper. Depending on the number of compactified dimensions, the relevant vector bundle underlying the geometric setup changes \cite{hull2007generalised} for example to
\begin{align}
    & TM \oplus \Lambda^2 T^*M\, , \nonumber\\
    & TM \oplus \Lambda^2 T^*M \oplus \Lambda^5 T^*M\, , \nonumber\\
    & TM \oplus \Lambda^2 T^*M \oplus \Lambda^5 T^*M \oplus \left( T^*M \otimes \Lambda^7 T^*M \right) \, , 
\label{physicsbundles}
\end{align}
which can all be equipped with certain exceptional Courant brackets \cite{pacheco2008m} realizing the symmetries in stringy geometry and M theories. For the simplest case $TM \oplus \Lambda^2 T^*M$, which can be endowed with a higher Courant algebroid structure \cite{bi2011higher}, there are crucial advancements. For instance, the analogous relation between Bianchi identities and consistency condition for a threebrane sigma model is investigated in \cite{chatzistavrakidis2019fluxes}. In this case, the number of compactified dimensions is 4, and the resulting duality group is $E_{4(4)}$ isomorphic to $SL(5)$. Mimicking the ideas of Poisson-Lie T-duality for this case, recently $SL(5)$ exceptional Drinfel'd algebra is introduced \cite{sakatani2020u, malek2020poisson}. Their extensions to more complicated vector bundles are further studied in \cite{malek2021e6, blair2022generalised}. Analogous to DFT, Exceptional Field Theories (ExFT) require extra coordinates coming from such vector bundles to make the $E_{d(d)}$ symmetry manifest \cite{hohm2013exceptional}. These extra dimensions are physically realized by membrane winding modes or Kaluza-Klein charges, and they have more complicated forms than just the doubling by the dual cotangent bundle \cite{Berman:2011jh}. Similarly to the DFT case, in order to have a physically consistent theory one has to impose a constraint. Under this constraint, the geometrical structures in ExFT are described within the framework of exceptional generalized geometry \cite{pacheco2008m, bugden2021g, hulik2024algebroids}. To further this progress, a closer look on the geometry on direct sum vector bundles equipped with a rather general form of an algebroid structure is of great importance. Such a geometry should extend the triple relation T-duality/DFT/Courant algebroids in the realm of U-duality and ExFT. With such motivations, in \cite{drinfeldpaper}, we defined a more general notion of bialgebroid for vector bundles which are not dual in the usual sense. Particularly, this work extends Lie bialgebroids \cite{mackenzie1994lie}, their Drinfel'd doubles \cite{liu1997manin} and matched pairs of Leibniz algebroids \cite{ibanez2001matched}.

Our study here achieves two distinct generalizations: On the one hand, we extend our notion of bialgebroids to \textit{proto bialgebroids} where both $H$- and $R$-twists are present. On the other hand, we generalize proto Lie bialgebroids of \cite{roytenberg2002quasi} to a pair of vector bundles that are not necessarily dual. The outcome of both of these generalizations do coincide so that novel construction we introduce in this paper can be considered as the up-right corner of the following diagram: 
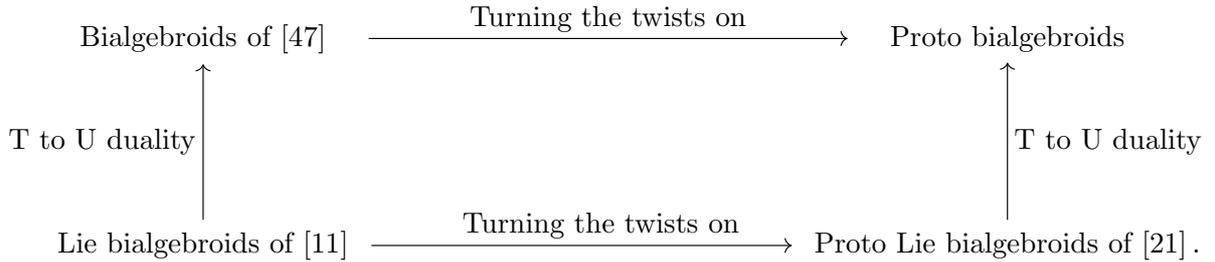
\[\begin{tikzcd}
	{\quad \mbox{Bialgebroids of \cite{drinfeldpaper}} \quad} &&&&&& {\quad \mbox{Proto bialgebroids} \quad} \\
	\\
	\\
	{\mbox{\ Lie bialgebroids of \cite{mackenzie1994lie}} \ } &&&&&& {\ \mbox{Proto Lie bialgebroids of \cite{roytenberg2002quasi}} \, . \ }
	\arrow["{\mbox{Turning the twists on}}", from=1-1, to=1-7]
	\arrow["{\mbox{T to U duality}}", from=4-1, to=1-1]
	\arrow["{\ \ \ \ \ \ \ \ \ \ \ \ \mbox{Turning the twists on} \ \ \ \ \ \ \ \ }", from=4-1, to=4-7]
	\arrow["{\mbox{T to U duality}}"', from=4-7, to=1-7]
\end{tikzcd}\]
Our strategy is to perform an analysis of certain algebroid axioms that are frequently used in the literature similarly to our previous work \cite{drinfeldpaper}, but now in the presence of both twists. In \cite{drinfeldpaper} we introduce the notion of \textit{calculus} on algebroids as a triplet of maps $(\mathcal{L}, \iota, d)$ which generalizes the usual Cartan calculus of Lie derivative, interior product and exterior derivative acting on differential forms. Equipped with two calculi, we are interested in the general form of a \textit{Roytenberg bracket}
\begin{equation}
    [U + \omega, V + \eta]_E = [U, V]_A + \tilde{\mathcal{L}}_{\omega} V - \tilde{\mathcal{L}}_{\eta} U + \tilde{d} \tilde{\iota}_{\eta} U + R(\omega, \eta) \oplus [\omega, \eta]_Z + \mathcal{L}_U \eta - \mathcal{L}_V \omega + d \iota_V \omega + H(U, V) \, ,
\end{equation}
on an arbitrary vector bundle of the form 
\begin{equation}
    E = A \oplus Z \, ,
\end{equation}
where $A$ and $Z$ are vector bundles which are not necessarily dual and of arbitrary ranks. We analyze each algebroid axiom individually and derive the ``twisted compatibility conditions'' on these calculus elements in order for the above bracket to satisfy desired properties. We explicitly investigate the modifications arising due to the presence of both $H$- and $R$-twists in our calculus framework. In particular we examine the following algebroid axioms: right- and left-Leibniz rules, Jacobi identity, symmetric part decomposition for the bracket and twists, metric invariance property and certain morphisms of brackets. Relaxation of algebroid structures usually has implications at the physical level since algebroid properties correspond to crucial consistency conditions or Bianchi identities of sigma models \cite{ikeda2003chern, blumenhagen2012bianchi} as we have mentioned. Hence our general and individual analysis of algebroid axioms has importance for the mathematics of exceptional field theories and U-duality, where the relevant vector bundles are indeed of the form~$E = A \oplus Z$. 

In light of our observations about algebroid axioms and twisted compatibility conditions, we introduce the notion of proto bialgebroids extending our notion of bialgebroids \cite{drinfeldpaper} in the presence of twists. We loosely use the term proto bialgebroid without any adjectives like we did for bialgebroids in our previous work \cite{drinfeldpaper} as we examine the axioms individually. These constructions also generalize the notion of proto Lie bialgebroids of Roytenberg \cite{roytenberg2002quasi}, as indicated by the right arrow in the diagram. We explicitly prove that this is indeed the case, where we present this generalization without using the supermanifold formalism. Hence our constructions can be considered as ``bosonic'' generalizations in the spirit of \cite{chatzistavrakidis2015sigma} which we also discuss in details. We define the notion of Drinfel'd double of an arbitrary proto bialgebroid. Our main theorem in Section \ref{s6} states that the Drinfel'd double of a proto bialgebroid satisfying twisted compatibility conditions derived in Section \ref{s5} for a desired set of axioms satisfies the same set of axioms. This result extends the fact that the Drinfel'd doubles of both Lie bialgebroids and proto Lie bialgebroids are Courant algebroids. Furthermore, we define and comment on the special cases in the presence of only one twist, generalizing quasi Lie and Lie quasi bialgebroids \cite{roytenberg2002quasi}. We present several examples for all these cases from physics and mathematics literatures. We also introduce the notion of proto metric-Bourbaki bialgebroid which satisfy all the twisted compatibility conditions; extending our previous works \cite{ccatal2022pre, drinfeldpaper}. 

We continue with describing a certain procedure to construct new calculi and twists from a given initial bracket, which depends on a choice of an automorphism of $E = A \oplus Z$, usually referred to as a twist matrix in the physics literature. This technique is often used to find out the missing terms of the Roytenberg bracket in terms of certain maps including $B$- and $\beta$-transformations. For instance, in \cite{halmagyi2009non} a bivector is used to construct the full Roytenberg bracket starting from the twisted Dorfman bracket. In \cite{drinfeldpaper}, we show that this construction can be considered as a reconstruction of triangular Lie bialgebroids of \cite{mackenzie1994lie}. There, we extend this triangularity notion in the realm of higher Courant algebroids \cite{bi2011higher} and Nambu-Poisson structures \cite{nambu1973generalized, takhtajan1994foundation}. Here, we give the most general form of such a twist automorphism procedure between two Roytenberg brackets. We explicitly evaluate the resulting calculi and twists yielded by a twist automorphism in this case together with the case that the initial choice of the bracket is given by a Dorfman-like bracket as usually done in the literature. We then focus on important special cases extending the usual notions of $B$- and $\beta$-transformations for arbitrary $E = A \oplus Z$.

The organization of the paper is as follows: In Section \ref{s2}, we outline the notation and conventions used throughout the paper. In the prelude, Section \ref{s3}, we summarize the basics of Lie bialgebroids and proto Lie bialgebroids together with their Drinfel’d doubles, where these structures are constructed on two dual vector bundles. Our main aim is to extend proto Lie bialgebroids for vector bundles which are not dual in the usual sense via an analysis of algebroid properties. To this end, we finish the prelude with a list of algebroid axioms that we use in our analysis. Section \ref{s4} is a summary of our earlier work \cite{drinfeldpaper} where we study the mentioned algebroid axioms on the direct sums of arbitrary vector bundles to define the notion of bialgebroid which is a generalization of Lie bialgebroids. Here following \cite{drinfeldpaper}, we introduce the calculus framework on algebroids. Moreover we examine each algebroid axiom separately and give explicit compatibility conditions. This analysis lays the groundwork for Section \ref{s5} where we present our novel and main calculations: We study all the algebroid axioms in the presence of both $H$- and $R$-twists. Here we follow the same steps as in Section \ref{s4} and present twisted compatibility conditions as modifications coming from the twists for each axiom. Next, in Section \ref{s6}, we present the main definitions and results in light of our observations from the previous one. Here, we define the notion of proto bialgebroid which is a natural generalization of bialgebroids we defined \cite{drinfeldpaper} in the presence of twists. Then we explicitly prove that this definition extends the proto Lie bialgebroids \cite{roytenberg2002quasi} defined for a pair of dual vector bundles; completing our diagram. We also look at certain subcases when either a single or both of the twists are absent. Different from its original formulation in the language of supermanifolds, here we present a bosonic definition similarly to the one given in \cite{chatzistavrakidis2015sigma} which is more appropriate for physical applications. We also state our main theorems regarding the Drinfel'd doubles of proto bialgebroids, and in particular proto metric-Bourbaki bialgebroids. Sections \ref{s5} and \ref{s6} consist of the original and main contributions of the paper. Afterwards in Section \ref{s7} we study effects of vector bundle automorphisms on given algebroid structures such as brackets, calculi and twists. To keep the discussion as general as we can, we present a twist automorphism procedure relating two Roytenberg brackets, and then focus on implications of particular subcases for such a twist procedure. Section \ref{s8} consists of examples of various proto bialgebroids from both physics and mathematics literatures. These examples include higher Courant, conformal Courant, $AV$-Courant, $E$-Courant and Atiyah algebroids together with $B_n$-generalized geometry. In Section \ref{s10}, we make some concluding remarks and briefly discuss some future works that we plan. Finally in Appendix, we summarize the basics of Cartan calculus on (almost-)Lie algebroids. Moreover, we relate them to our own calculus framework and use some of these results in the proof of Section~\ref{s6}.

%%%%%%%%%%%%%%%%%%%%%%%%%%%%%

\section{Notation and Conventions}
\label{s2}

In this section, we introduce the notation of the paper and set our conventions. As this paper is a natural follow up of our previous work \cite{drinfeldpaper}, we use the same notation and conventions.

Every structure in this paper is assumed to be smooth, and we always consider a connected, paracompact, Hausdorff and orientable manifold $M$. All vector bundles are over the same base manifold $M$ with a specified projection map, and they are assumed to be real and of finite-rank. The tangent and cotangent bundles are respectively denoted by $TM$ and $T^*M$, and the ring of (real-valued) smooth functions is denoted by $C^{\infty}M$. Sections of the tangent bundle are vector fields and they act on smooth functions as derivations, so that for a vector field $V$ its action on a smooth function $f$ is denoted by $V(f)$. Exterior power of a vector bundle is denoted by $\Lambda$; in particular the sections of $\Lambda^p T^*M$ are (exterior differential) $p$-forms. We only consider vector bundle morphisms that are over $\text{id}_M$, where we denote the identity map with $\text{id}$. Consequently, a vector bundle morphism is denoted only by a single map $\Phi: E_1 \to E_2$. Vector bundle morphisms induce $C^{\infty}M$-linear maps on the sections; in fact every map between two vector bundles are to be understood as section-wise. Particularly, when we consider vector bundle morphisms with the same range and domain, say $E$, we call the invertible ones as automorphisms and denote their set by $\text{Aut}(E)$. The composition of two maps is just juxtaposition, so that we do not use any symbol for it. Moreover, with an abuse of notation, we directly write $u, v, w \in E$ for sections $u, v, w$ of a vector bundle $E$.

One can completely analogously generalize the notions of tensors, vector fields,  $p$-forms, interior product, frames and coframes on an arbitrary vector bundle $E$ by replacing the tangent bundle with $E$ and the cotangent bundle with $E^*$ (see for a detailed exposition \cite{dereli2021metric}). In particular we call $C^{\infty}$-bilinear symmetric maps $g_E: E \times E \to \mathbb{E}$ as metrics on $E$ taking values in $\mathbb{E}$, and we do not care about non-degeneracy. Sections of an arbitrary vector bundle $E$ do not act as derivations on smooth functions, so for various structures including connections and brackets, one needs to introduce an additional ingredient which circumvents this problem. This additional item is the \textit{anchor}, which is a vector bundle morphism $\rho_E: E \to TM$. In this paper we will consider only anchored vector bundles, \textit{i.e.}, vector bundles are assumed to be equipped with an anchor. Given two vector bundles $E$ and $A$, one can define the  of $E$-connection on $A$ as an $\mathbb{R}$-bilinear map $\nabla: E \times A \to A$ satisfying \cite{fernandes2002lie}
\begin{align}
    \nabla_u (f U) &= f \nabla_u U + \rho_E(u)(f) U \, , \nonumber\\
    \nabla_{f u} U &= f \nabla_u U \, ,
\end{align}
for all $u \in E, U \in A, f \in C^{\infty}M$.

Throughout the paper, we will be interested in vector bundles which can be written as a direct sum, for instance $E = A \oplus Z$. We denote the projection maps to $A$ and $Z$ with $\text{pr}_A$ and $\text{pr}_Z$, respectively. When we want to make the $A$- and $Z$-parts of a section transparent, we use $\oplus$ notation. For example, for a section $u$ of $E$, $u = U \oplus \omega$ means that $U$ and $\omega$ are sections of $A$ and $Z$, respectively. Keeping in mind that we want to generalize the notion of Courant algebroids which are typically of the form $TM \oplus T^*M$, we use capital Latin letters $U, V, W$ for the sections of $A$, and lowercase Greek letters $\omega, \eta, \mu$ for the sections of $Z$.

In our formulation we try to be as general as we can, therefore we have lots of different vector bundles with several maps between them. We name these vector bundles in a particular way which we hope makes keeping track of the domains and ranges easier. We use different fonts like $\mathbb{A}, \mathcal{A}, \mathscr{A}$ (resp. $\mathbb{Z}, \mathcal{Z}, \mathscr{Z}$) to indicate that they are somewhat related to the vector bundle $A$ (resp. $Z$). In particular, the symbol $\mathbb{R}$ does not denote any vector bundle, which is just the set of real numbers. We also sometimes behave like that the ring of smooth functions $C^{\infty}M$ is a vector bundle in order to prevent writing something like $\Lambda^0 T^*M$. 

In this paper, we are interested in various generalizations of Cartan calculus operators; namely Lie derivative $\mathcal{L}$, interior product $\iota$ and exterior derivative $d$. We summarize their basic properties in Appendix on (almost-)Lie algebroids, but in several sections the same letters usually denote their abstract generalizations which we introduce in Section \ref{s4} and Section \ref{s5}.

%%%%%%%%%%%%%%%%%%%%%%%%%%%%%%

\section{Prelude: Lie Bialgebroids to Proto Lie Bialgebroids}
\label{s3}

In this section, we briefly review the Lie bialgebroid structures \cite{mackenzie1994lie}, and summarize the basics of proto Lie bialgebroids of Roytenberg \cite{roytenberg2002quasi} which are defined for a pair of vector bundles that are dual. Moreover we discuss the fact that the Drinfel'd doubles of both of them yield a Courant algebroid \cite{liu1997manin, roytenberg2002quasi} which we also define here. In the consecutive sections, our main aim  will be to extend these notions in a more general setting. Our guiding principle when extending them to direct sums of arbitrary vector bundles that are not dual in the usual sense will be to study certain algebroid axioms which come from generalizing the properties of Courant algebroids. We present the general forms of these algebroid axioms at the end of this section which lay the groundwork for analysis of the following two sections.

A pair of Lie algebra structures on dual vector spaces $(\mathfrak{g}, \mathfrak{g}^*)$ is called a Lie bialgebra if they are compatible in a certain sense \cite{drinfeld1986quantum}. For a Lie bialgebra, there is a unique Lie algebra structure on $\mathfrak{d} = \mathfrak{g} \oplus \mathfrak{g}^*$ satisfying certain properties, called its Drinfel'd double. This  can be extended to the Lie algebroid level, where a triplet $(A, \rho_A, [\cdot,\cdot]_A)$ is called a Lie algebroid if $(A, [\cdot,\cdot]_A)$ is a Lie algebra over $\mathbb{R}$ further satisfying the right-Leibniz rule:
\begin{equation}
    [U, f V]_A = f [U, V]_A + \rho_A(U)(f) V \, .
\end{equation}
A Lie bialgebroid is then a pair of dual Lie algebroids
$(A, A^*)$, with the compatibility condition \cite{mackenzie1994lie} 
\begin{equation}
    d^*[U, V]_A = \mathcal{L}_U d^* V - \mathcal{L}_V d^* U \, ,
\label{compatibilityLie}
\end{equation}
where $d^*$ is the exterior derivative induced by the bracket $[\cdot,\cdot]_{A^*}$ and $\mathcal{L}$ is the Lie derivative induced by the bracket $[\cdot,\cdot]_A$ (for details see Appendix). Equivalently, $d^*$ is a derivation of the Schouten-Nijenhuis bracket constructed from the bracket of $A$ \cite{kosmann1995exact}. For a (almost-)Lie algebroid $A$, the \textit{Schouten-Nijenhuis bracket} $[\cdot,\cdot]_{\text{SN},A}: \Lambda^p A \times \Lambda^q A \to \Lambda^{p + q - 1} A$ is defined as an extension of the bracket $[\cdot,\cdot]_A$:
\begin{equation}
    [U_1 \wedge \ldots \wedge U_p, V_1 \wedge \ldots \wedge V_q]_{\text{SN},A} := \sum_{i = 1}^p \sum_{j = 1}^q  (-1)^{i + j} [U_i, V_j]_A \wedge U_1 \wedge \ldots \wedge \check{U_i} \wedge \ldots \wedge U_p \wedge V_1 \wedge \ldots \wedge \check{V_j} \wedge \ldots \wedge V_q \, ,
\label{schouten}
\end{equation}
with $[U, f]_{\text{SN},A} = - [f, U]_{\text{SN},A} = \rho_A(U)(f)$ for a smooth function $f$. Here, $\check{U_i}$ indicates that $U_i$ is excluded. This bracket makes the multivector fields of $A$ a Gerstenhaber algebra, when $A$ is a Lie algebroid. 

The Drinfel'd double of a Lie bialgebroid $(A, A^*)$, which is the unique algebroid structure on $E = A \oplus A^*$ such that both $A$ and $A^*$ are Lie subalgebroids and the canonical pairing is invariant under the adjoint action on $E$, is a Courant algebroid instead of a Lie algebroid \cite{liu1997manin}. Courant algebroids are defined as a quadruplet $(E, \rho_E, [\cdot,\cdot]_E, g_E)$, where $g_E: E \times E \to C^{\infty}M$ is a non-degenerate metric and the following are satisfied:
\begin{align}
    \rho_E(u)(g_E(v, w)) &= g_E([u, v]_E, w) + g_E(v, [u, w]_E) \, , \nonumber\\
    [u, v]_E + [v, u]_E &= g_E^{-1} D_E g_E(u, v) \, , \nonumber\\
    [u, [v, w]_E]_E &= [[u, v]_E, w]_E + [v, [u, w]_E]_E \, .
\end{align}
In the symmetric part, the map $D_E: C^{\infty}M \to E^*$ is defined as $D_E f := \rho^t d$ with $d$ being the usual exterior derivative and $^t$ denotes the transpose of a linear map. Courant algebroids satisfy the left-Leibniz rule
\begin{equation}
    [f u, v]_E = f [u, v]_E - \rho_E(v)(f) u + g_E(u, v) g_E^{-1}(D_E f) \, .
\end{equation}
One of the fundamental results about Courant algebroids is the \v{S}evera classification theorem \cite{vsevera2017letters} of exact Courant algebroids which fit into the following exact sequence 
\begin{equation}
    0 \xrightarrow{\quad} T^*M \xrightarrow{\ \ g_E^{-1} \rho_E^t \ \ } E \xrightarrow{\ \ \rho_E \ \ } TM \xrightarrow{\quad} 0 \, .
\label{exactcourant}
\end{equation}
According to this theorem, any exact Courant algebroid is isomorphic to the generalized tangent bundle $TM \oplus T^*M$ equipped with the $H$-twisted version of the Dorfman bracket \cite{dorfman1987dirac}
\begin{equation}
    [U + \omega, V + \eta]_{\text{Dor}} := [U, V]_{\text{Lie}} \oplus \mathcal{L}_U \eta - \mathcal{L}_V \omega + d \iota_V \omega \, ,
\end{equation}
namely for a closed 3-form $H$:
\begin{equation}
    [U + \omega, V + \eta]_{\text{Dor}}^H := [U, V]_{\text{Lie}} \oplus \mathcal{L}_U \eta - \mathcal{L}_V \omega + d \iota_V \omega + H(U, V) \, .
\end{equation}
Due to the necessary form of the algebroid structure given by the twisted Dorfman bracket, exact Courant algebroids only sustain the geometric $f$ and $H$ fluxes. The anti-symmetrization of the Dorfman bracket is known as the Courant bracket \cite{courant1990dirac}, and Courant algebroid definition evolved through time \cite{liu1997manin, roytenberg1999courant, uchino2002remarks}. When the Jacobi identity is relaxed, then the structure is called a metric algebroid \cite{vaisman2012geometry}.

Lie algebroids and bialgebroids can be also neatly explained by using supermanifolds formalism \cite{vaintrob1997lie, roytenberg2002quasi, roytenberg1999courant} (for a detailed exposition of supermanifold formalism, we refer to \cite{manin2013gauge}, see also \cite{cattaneo2011introduction}). The equivalence of the Roytenberg's construction that we give below to the one given by Mackenzie and Xu is established by Voronov in \cite{voronov2012q}. Given a vector bundle $A$, its parity shifted version $\Pi A$ is a supermanifold whose sheaf of functions is identified with sections of the associated exterior bundle of the dual vector bundle, that is $\Lambda A^*$. Similarly its parity shifted dual $\Pi A^*$ is also a supermanifold, and both $\Pi A$ and $\Pi A^*$ sit inside the even symplectic supermanifold $(T^*\Pi A, \{ \cdot, \cdot \})$ equipped with the canonical symplectic form whose Poisson bracket is denoted by $\{ \cdot, \cdot \}$. Moreover there is a bi-grading on superfields on $T^* \Pi A$ which separately counts the Grassmann degrees of $\Pi A$ and $\Pi A^*$ inside~$T^* \Pi A$. 

In this setting, a Lie algebroid structure on $A$ is equivalent to a degree-1 homological vector field on $\Pi A$, \textit{i.e.}, a derivation on $\Lambda A^*$ which increases the degrees by one and squares to zero. This is equivalent to a function $\mu$ on $T^*\Pi A$ of bi-degree $(1, 2)$ satisfying $\{ \mu, \mu \} = 0$. Similarly, a Lie algebroid structure on $A^*$ is encoded by a function $\gamma$ of bi-degree $(2,1)$ satisfying $\{ \gamma, \gamma \} = 0$. Then a Lie bialgebroid structure on $(A,A^*)$ is defined as a pair of Lie algebroids furthermore requiring $\{ \mu, \gamma \} = 0$ which is nothing but the compatibility condition (\ref{compatibilityLie}). This can be repackaged into a superfield $\theta = \mu + \gamma$ on $T^*\Pi A$ of total degree 3 satisfying the master equation
\begin{equation}
    \{ \theta, \theta \} = 0 \, .
\label{masterequation}
\end{equation}
In particular, the Courant algebroid structure on the double $E = A \oplus A^*$ is given by the derived bracket \cite{Kosmann_Schwarzbach_2004, kosmann2011poisson}:
\begin{align}
    [u, v]_E &= \{ \{ u, \theta \}, v \} \, , \nonumber\\
    \rho_E(u)(f) &= \{ \{ u, \theta \}, f \} \, , \nonumber\\
    g_E(u, v) &= \{ u, v \} \, ,
\end{align}
where sections of $A$ and $A^*$ are identified with linear superfields on $\Pi A^*$ and $\Pi A$, respectively. 

The Lie bialgebroid structure defined above can easily be generalized by considering the most general degree 3 superfield $\Theta$ on $T^* \Pi A$. It splits according to $\Theta = \mu + \gamma + \phi + \psi$, where $\phi$ and $\psi$ are functions of bi-degree $(0, 3)$ and $(3,0)$, respectively. Consequently $\phi$ and $\psi$ are 3-forms on $A$ and $A^*$, respectively. If $\Theta$ obeys the master equation (\ref{masterequation}), then $(A, A^*)$ is called a proto Lie bialgebroid. Remarkably, the Drinfel'd double $E = A \oplus A^*$ still remains a Courant algebroid \cite{roytenberg2002quasi}. Splitting the master equation according to bi-grading, one gets
\begin{align}
    \tfrac{1}{2} \{ \mu, \mu \} + \{ \gamma, \phi \} &= 0 \, , \label{roytenberg1} \\
    \tfrac{1}{2} \{ \gamma, \gamma \} + \{ \mu, \psi \} &= 0 \, , \label{roytenberg2} \\
    \{ \mu, \gamma \} + \{ \phi, \psi \} &= 0 \, ,  \label{roytenberg3} \\
    \{ \mu, \phi \} &= 0 \, , \label{roytenberg4} \\
    \{ \gamma, \psi \} &= 0 \, . \label{roytenberg5}
\end{align}
When $\phi$ (resp. $\psi$) is zero, the structure is termed a Lie quasi bialgebroid (resp. quasi Lie bialgebroid), where the fourth equation (resp. fifth equation) holds trivially. For both of these cases, third equation above becomes the usual compatibility condition $\{ \mu, \gamma \} = 0$. All of these are twisted generalizations of Lie bialgebroids.

The Courant algebroid corresponding to the Drinfel'd double of a proto Lie bialgebroid conveniently describes both geometric and non-geometric fluxes associated with sting theory compactifications. In particular for $A = TM$ on a suitably chosen frame, the form of the Roytenberg bracket (\ref{roytenbergbracket}) agrees with the fluxes (\ref{km}) appearing in the works of Kaloper and Myers \cite{kaloper1999dd}. In physical constructions one generally does not have the full Roytenberg bracket. This is because when working with the generalized tangent bundle $TM \oplus T^*M$, one only has the action of vectors on vectors and on forms through the Lie bracket. Without any additional structure such as a Poisson bivector, the forms has no canonical action on themselves or vectors. However, by supplying appropriate additional structures, such Roytenberg brackets can generally be obtained from the Dorfman bracket on the generalized tangent bundle as in \cite{halmagyi2009non}. More general forms of twists matrices can be found in several papers including for example \cite{coimbra2011supergravity, chatzistavrakidis2018double, tomas2020generalized, coimbra2014e_d, baraglia2012leibniz}, as we will come back in Section \ref{s7}.

Since the full Roytenberg bracket on the Drinfel'd double of a proto Lie bialgebroid includes both $H$- and $R$-twists, it can incorporate all four of the fluxes: $f, H, Q, R$. On the other hand, the Drinfel'd double of a Lie bialgebroid only sustains $f$ and $Q$ fluxes as the summands are subalgebroids of the double. The fluxes associated with this Roytenberg bracket can be physically realized in two ways in general. At the level of the string sigma model, these fluxes are related to the Wess-Zumino-Witten terms coming from the bulk uplift of the Lagrangian \cite{klimvcík2002wzw}. Equivalently these fluxes appear in the deformed current algebra associated with the Hamiltonian of this sigma model \cite{halmagyi2009non}. The relation between non-geometric fluxes and deformed current algebras, including the ones which do not have Lagrangian descriptions, are studied in detail in \cite{osten2020current}. A similar analysis is used to classify all possible fluxes coming from appropriate twist matrices on the generalized tangent bundle which is utilized as a solution generating technique in supergravity \cite{borsato2021algebraic}.

Ultimately, some of these ideas can be extended to non-dual algebroids $A$, $Z$, and their Drinfel'd double $A \oplus Z$ in the context of membrane models, and the geometry on such vector bundles will be the main subject of this paper. Noticeably, the Roytenberg algebras coming from a bivector twist is generalized to higher Roytenberg algebras on $TM \oplus \Lambda^p T^*M$ using a twist induced by $(p + 1)$-vectors \cite{jurvco2013p, drinfeldpaper}. In the case that the multivector is a Nambu-Poisson structure, the $R$-twist vanishes. Moreover the membrane action associated with this model is studied and fluxes are again related to the generalized Wess-Zumino-Witten terms appearing on the bulk uplift. Nambu-Poisson structures play also an important role in sigma models of membranes \cite{bagger2007modeling}. In our earlier study we also showed that twisting higher Dorfman bracket using a Nambu-Poisson structure yields a pair of dual calculi compatible in the sense that will be explained in Section \ref{s4} \cite{drinfeldpaper}. These studies are important, because they lay groundwork for the study of more general bundles (\ref{physicsbundles}) associated with U-dualities all of which can be endowed with exceptional Courant brackets \cite{pacheco2008m}. Particularly for $SL(5)$ M theory, the bundle encapsulating the generalized symmetries coming from 4 dimensional torus compactifications is $TM \oplus \Lambda^2 T^*M$. In this case, $SL(5)$ exceptional Drinfel'd algebra has been constructed using a twist induced by Nambu-Poisson trivector \cite{sakatani2020u, malek2020poisson}. An initial analysis shows that our dual calculus framework is suitable to study this exceptional Drinfel'd algebra as we obtain the non-constant fluxes in the construction of this algebra. 

As we mentioned above, the Drinfel'd double of a proto Lie bialgebroid is a Courant algebroid, which plays an important role in T-duality. We also commented on a need for generalizations of such doubled algebroid structures on more arbitrary vector bundles. Relaxation of algebroid structures often have a deep physical meaning as we highlighted above. Hence, the analysis of rather general forms of algebroid properties is of great importance for the mathematics of U-duality. Such an analysis will be the main focus in the next sections based on our previous work \cite{drinfeldpaper}. Therefore, we finish the prelude with a summary of general forms of algebroid axioms that are frequently used in the literature and will be analyzed in the upcoming sections.

An algebroid $(E, \rho_E, [\cdot,\cdot]_E)$ is said to satisfy the right-Leibniz rule if
\begin{equation}
    [u, f v]_E = f [u, v]_E + \rho_E(u)(f) v \, .
\label{rightleibnizrule}
\end{equation}
Such algebroids are sometimes called almost-Leibniz algebroids. Similarly it is said to satisfy the left-Leibniz rule if for a map of the form $L_E: C^{\infty}M \times E \times E \to E$ with appropriate $C^{\infty}M$-linearity properties, which is called the locality operator
\begin{equation}
    [f u, v]_E = f [u, v]_E - \rho_E(v)(f) u + L_E(f, u, v) \, ,
\label{leftleibnizrule}
\end{equation}
and an algebroid satisfying both Leibniz rules is called a local almost-Leibniz algebroid. The almost prefix is replaced with the pre prefix when the anchor is a morphism of brackets in the sense that
\begin{equation}
    \rho_E([u, v]_E) = [\rho_E(u), \rho_E(v)]_{\text{Lie}} \, .
\end{equation}
Note that the use of pre and almost prefixes might differ from paper to paper. Moreover the term proto bialgebroid (or protobialgebroid or proto-bialgebroid, see for example \cite{chatzistavrakidis2015sigma, kosmann2011poisson}) is often used as proto Lie bialgebroids as defined in this section. However, we reserve the term proto bialgebroid for the generalization of proto Lie bialgebroids that we will introduce in Section \ref{s6}.  

An algebroid is said to satisfy the Jacobi identity if
\begin{equation}
    [u, [v, w]_E]_E = [[u, v]_E, w]_E + [v, [u, w]_E]_E \, ,
\label{jacobiidentity}
\end{equation}
or equivalently the bracket's Jacobiator
\begin{equation}
    \mathcal{J}_E(u, v, w) := [u, [v, w]_E]_E - [[u, v]_E, w]_E - [v, [u, w]_E]_E \, 
\end{equation}
vanishes. Algebroids satisfying the right-Leibniz rule and the Jacobi identity are called Leibniz algebroids. Moreover, Jacobi identity together with the right-Leibniz rule implies that the anchor is a morphism of brackets. Many algebroids come with a bracket whose symmetric part can be decomposed as
\begin{equation}
    [u, v]_E + [v, u]_E = \mathbb{D}_E g_E(u, v) \, ,
\label{symmetricpart}
\end{equation}
where $g_E: E \times E \to \mathbb{E}$ is an $\mathbb{E}$-valued (possibly degenerate) metric on $E$ and $\mathbb{D}_E: \mathbb{E} \to E$ is a first-order differential operator. When the symmetric part vanishes, it is the realm of Lie algebroids. For example an algebroid equipped with an anti-symmetric bracket satisfying the right-Leibniz rule is called an almost-Lie algebroid, and it is a Lie algebroid if it further satisfies the Jacobi identity. Lastly, the metric invariance property for such an $\mathbb{E}$-valued metric on $E$ can be written in the general form
\begin{equation}
    \mathbb{L}^E_u g_E(v, w) = g_E([u, v]_E, w) + g_E(v, [u, w]_E) \, ,
\label{metricinvariance}
\end{equation}
for a first-order differential operator $\mathbb{L}^E: E \times \mathbb{E} \to E$, which we refer as \textit{metric invariance operator}. A septet $(E, \rho_E, [\cdot,\cdot]_E, g_E, \mathbb{E}, \mathbb{D}_E, \mathbb{L}^E)$ is called a metric-Bourbaki algebroid \cite{ccatal2022pre} (or as we have noticed very recently; the same construction is termed Vinogradov algebroids in\cite{kotov2010generalizing}) if it satisfies all of these properties. These algebroids are natural generalizations of Courant algebroids for which we identify $\mathbb{E} = C^{\infty}M, \mathbb{D}_E = g_E^{-1} D_E$ and $\mathbb{L}^E = \rho_E$. Special cases of metric-Bourbaki algebroids include Lie, higher Courant \cite{bi2011higher}, conformal Courant \cite{baraglia2013conformal}, $AV$-Courant \cite{libland2011} and omni-Lie algebroids \cite{chen2010omni}.

%%%%%%%%%%%%%%%%%%%%%%%%%%%%%%

\section{Calculus on Algebroids and Bialgebroids}
\label{s4}

In this section, following our previous work \cite{drinfeldpaper}, we summarize the framework of calculus on algebroids associated to two vector bundles $A$ and $Z$ which are not necessarily dual and of arbitrary ranks. As opposed to (proto) Lie bialgebroids, the summands of $A \oplus Z$ being not dual is key for applications of ``doubled'' structures regarding exceptional geometries and U-duality. By examining the frequently used algebroid properties from the literature in their rather general form, we explicitly evaluate compatibility conditions. These conditions equip the pair $(A, Z)$ with a relaxed notion of bialgebroid, and their Drinfel'd double $E = A \oplus Z$ with an algebroid structure satisfying desired properties. It is important to note that in our analysis $A$ and $Z$ are at an equal footing, that is neither of them are special compared to one another. Therefore when we interchange the roles of them, we obtain an analogous set of structures and conditions which we loosely call as \textit{dual} structures or conditions. In order to find a suitable generalization for bialgebroids, here we introduce the notion of calculus on algebroids which will be the main framework throughout the paper.

In our calculations, two important algebroid properties that we analyze are the symmetric part decomposition of the bracket (\ref{symmetricpart}) and metric invariance property (\ref{metricinvariance}). We want to emphasize that studying these axioms are well-suited for physical applications and generally not considered in mathematical studies regarding direct sums of algebroid structures, such as the ones that we closely follow \cite{ibanez2001matched, mackenzie1994lie, roytenberg2002quasi}. The symmetric part decomposition will be especially important in the next section where we consider the same properties in the presence of twists. As the twists themselves might be asymmetric in general, we will assume a similar decomposition for the symmetric parts of the twists, which in particular will be relevant for the generalization of metric-Bourbaki bialgebroids.

A general algebroid structure on a direct sum $E = A \oplus Z$ is induced by the anchor
\begin{equation}
    \rho_E = \rho_A \oplus \rho_Z \, ,
\end{equation}
and a following type of bracket when both $A$ and $Z$ are subalgebroids of $E$ with the respective brackets $[\cdot,\cdot]_A$ and $[\cdot,\cdot]_Z$:
\begin{equation}
    [U + \omega, V + \eta]_E = [U, V]_A + \tilde{\mathcal{L}}_{\omega} V + \tilde{\mathcal{K}}_{\eta} U \oplus [\omega, \eta]_Z + \mathcal{L}_U \eta + \mathcal{K}_V \omega \, ,   
\label{untwistedbracket}
\end{equation}
for some first-order differential operators of the form 
\begin{align}
    &\mathcal{L}: A \times Z \to Z \, , \qquad \qquad \mathcal{K}: A \times Z \to Z \, \nonumber\\
    &\tilde{\mathcal{L}}: Z \times A \to A \, , \qquad \qquad \tilde{\mathcal{K}}: Z \times A \to A \, . \label{thesemaps}
\end{align}
This bracket can be decomposed into two Dorfman-like brackets
\begin{align}
    [U + \omega, V + \eta]_{\text{Dor}} &:= [U, V]_A \oplus \mathcal{L}_U \eta + \mathcal{K}_V \omega \, ,  \label{dorfmanbracket} \\ 
    [U + \omega, V + \eta]_{\widetilde{\text{Dor}}} &:= \tilde{\mathcal{L}}_\omega V + \tilde{\mathcal{K}}_\eta U \oplus [\omega,\eta]_Z \, , \label{dorfmantildebracket}
\end{align}
and we simply refer to them as Dorfman and tilde-Dorfman brackets.

Naturally the properties of the ``doubled bracket'' $[\cdot,\cdot]_E$ depend on the properties of the algebroid structures on $A$ and $Z$ together with the maps in (\ref{thesemaps}). In \cite{drinfeldpaper}, we observed that for the bracket (\ref{untwistedbracket}) to satisfy an axiom, the subalgebroids $A$ and $Z$ have to satisfy the same axiom. Furthermore, several compatibility conditions are derived for the algebroid properties which we summarize below. There, and also in this paper, we freely use the term bialgebroid for the pair of algebroids $(A, Z)$ satisfying a set of desired properties which are equipped with certain maps that we call calculus, satisfying the compatibility conditions for the desired properties. Then the algebroid structure on $E = A \oplus Z$ induced by the bracket (\ref{untwistedbracket}) is called the Drinfel'd double of the bialgebroid and as we show in \cite{drinfeldpaper}, the Drinfel'd double itself satisfies the same properties. As these compatibility conditions already come up as dual pairs for a bialgebroid $(A, Z)$, by definition we automatically have that $(Z, A)$ is also a bialgebroid. The notion of bialgebroid is a natural extension of Lie bialgebroids \cite{mackenzie1994lie, kosmann1995exact}, and also matched pairs of Leibniz algebroids \cite{ibanez2001matched}. A similar axiomatic analysis for metric and Courant algebroids together with their relaxed versions was done in \cite{mori2020doubled, mori2020more}. Our results from this and consecutive section recover some of the findings of these papers on doubled structures on algebroids. In particular, relaxing the compatibility condition for Lie bialgebroids, one gets a Drinfel'd double structure given by a metric algebroid \cite{mori2020doubled}. In these studies they relate the metric algebroid structures with the $C$-bracket encoding the gauge symmetries of DFT and discuss an ``algebraic origin'' of the strong constraint. In this sense it would be important to study the compatibility conditions as we do here since a similar relaxation of these axioms could lead to a better understanding of algebroid structures underlying ExFTs and in particular U-duality. 

To find a suitable generalization of the Cartan calculus, we decompose the maps $\mathcal{K}$ and $\tilde{\mathcal{K}}$ as
\begin{equation}
    \mathcal{K}_V \omega = - \mathcal{L}_V \omega + d \iota_V \omega \, , \qquad \qquad \qquad \tilde{\mathcal{K}}_{\eta} U = - \tilde{\mathcal{L}}_{\eta} U + \tilde{d} \tilde{\iota}_{\eta} U \, ,
\end{equation}
for arbitrary vector bundles $\mathcal{A}$ and $\mathcal{Z}$, and maps
\begin{align}
    &\iota: A \times Z \to \mathcal{Z} \, , \qquad \qquad d: \mathcal{Z} \to Z \, , \nonumber\\
    &\tilde{\iota}: Z \times A \to \mathcal{A} \, , \qquad \qquad \tilde{d}: \mathcal{A} \to A \, .
\end{align}
As for the usual interior product, we assume that the map $\iota$ is $C^{\infty}M$-bilinear. This decomposition is motivated by the fact that many algebroid structures have a bracket whose symmetric part can be decomposed in a particular way \cite{bugden2021g, ccatal2022pre} as in Equation (\ref{symmetricpart}). In order for the bracket (\ref{untwistedbracket}) to satisfy right- (\ref{rightleibnizrule}) and left-Leibniz rules (\ref{leftleibnizrule}), one may observe that the following \textit{linearity conditions} should be satisfied \cite{drinfeldpaper}
\begin{align}
    \Delta^{(1)}_{\mathcal{L}}(f, V, \omega) &= \Delta_d(f, \iota_V \omega) \, , \qquad \qquad \qquad \Delta^{(2)}_{\mathcal{L}}(f, V, \omega) = \rho_A(V)(f) \omega \, , \nonumber\\
    \Delta^{(1)}_{\iota}(f, V, \omega) &= 0 \, , \qquad \qquad \qquad \qquad \qquad \ \Delta^{(2)}_{\iota}(f, V, \omega) = 0 \, .
\label{linearityconditions}
\end{align}
Here symbol map $\Delta_\Phi$ of an arbitrary map $\Phi$ measures the non-tensoriality of $\Phi$, \textit{i.e.}, 
\begin{equation}
    \Delta_\Phi (f, u) := \Phi(f u) - f \Phi(u) \, .
\end{equation}
If the map $\Phi$ has a product domain, then we furnish $\Delta_\Phi$ with an upper index $(j)$ indicating the non-tensoriality in its $j$-th entry. For the Jacobi identity to hold for $E = A \oplus Z$, the Jacobi identities of $A$ and $Z$ should be satisfied:
\begin{equation}
    \mathcal{J}_A = 0 \, , \qquad \qquad \qquad \mathcal{J}_Z = 0 \, .
\end{equation}
Moreover, the following \textit{calculus conditions} should be satisfied as well\cite{drinfeldpaper}:
\begin{align}
    \mathcal{L}_U \mathcal{L}_V \mu - \mathcal{L}_V \mathcal{L}_U \mu - \mathcal{L}_{[U, V]_A} \mu &= 0 \, , \nonumber\\
    \mathcal{L}_U d \iota_W \eta - d \iota_{[U, W]_A} \eta - d \iota_W \mathcal{L}_U \eta &= 0 \, , \nonumber\\
        \mathcal{L}_W d \iota_V \omega - d \iota_W d \iota_V \omega &= 0 \, ,
\label{calculusconditions}
\end{align}
together with their duals for $(\tilde{\mathcal{L}}, \tilde{\iota}, \tilde{d})$. In \cite{drinfeldpaper}, we call a triplet of maps $(\mathcal{L}, \iota, d)$ as a \textit{calculus} on $Z$ induced by $A$ if the linearity and calculus conditions are satisfied. We sometimes refer to the second set of calculus $(\tilde{\mathcal{L}}, \tilde{\iota}, \tilde{d})$ as \textit{tilde-calculus}. In addition when it is clear from the context, we simply refer to $(\tilde{\mathcal{L}}, \tilde{\iota}, \tilde{d})$ and $(\mathcal{L}, \mathcal{K})$ as calculus elements. In fact we will still refer to these maps as calculus elements when we study their twistful generalizations in the following section. 

As we note and demonstrate in \cite{drinfeldpaper}, the notion of calculus provides a good book-keeping device useful for frame independent formulations suitable for physical applications. Note that Cartan calculus elements, Lie derivative, interior product and exterior derivative on a Lie algebroid (see Appendix for details) satisfy these properties, so that they form a calculus in the above sense. A calculus can be seen as a refinement of the notion of algebroid (Leibniz) representations \cite{ibanez2001matched}, where the latter in our notation is defined by a pair of maps $\mathcal{L}, \mathcal{K}: A \times Z \to Z$ satisfying
\begin{align}
     \mathcal{L}_{[U,V]_A} \omega &= \mathcal{L}_U \mathcal{L}_V \omega - \mathcal{L}_V \mathcal{L}_U \omega \, , \nonumber\\
     \mathcal{K}_{[U, V]_A} \omega &= \mathcal{K}_V \mathcal{K}_U \omega + \mathcal{L}_U \mathcal{K}_V \omega \, , \nonumber\\
     \mathcal{K}_U \mathcal{K}_V \omega &= - \mathcal{K}_U \mathcal{L}_V \omega \, ,
\label{leibnizrepresentation}
\end{align}
together with the $C^{\infty}M$-linearity properties
\begin{align}
    \mathcal{L}_U(f \omega) &= f \mathcal{L}_U \omega + \rho_A(U)(f) \omega \, , \nonumber\\
    \mathcal{K}_{f U} \omega &= f \mathcal{K}_U \omega \, .
\label{leibnizlinearity}
\end{align}
In this sense the notion of calculus is a special case of algebroid representations satisfying further properties.

The linearity and calculus conditions are the only conditions that do not mix two sets of calculus elements together in the analysis of algebroid axioms we are discussing. For example, Jacobi identity further imposes \textit{Jacobi compatibility conditions} \cite{drinfeldpaper}
\begin{align}
    \mathcal{D}^Z_{\mathcal{L}_U}(\eta, \mu) &= \mathcal{L}_{\tilde{\mathcal{K}}_{\eta} U} \mu + \mathcal{K}_{\tilde{\mathcal{K}}_{\mu} U} \eta \, , \label{comp1} \\
    \mathcal{L}_{\tilde{d} \tilde{\iota}_{\eta} U} \mu &= - [d \iota_U \eta, \mu]_Z \, , \label{comp2} \\
    d \iota_{\tilde{\mathcal{L}}_{\omega} W} \eta - d \iota_{\tilde{d} \tilde{\iota}_{\eta} W} \omega + d \iota_W [\omega, \eta]_Z &= \mathbb{D}_Z g_Z(d \iota_W \eta, \omega) - \mathbb{D}_Z g_Z(\mathcal{K}_W \omega, \eta) \, ,  \label{comp3} 
\end{align}
together with their duals, where we define the \textit{derivator} of a map $\Phi: E \to E$ as
\begin{equation}
    \mathcal{D}^E_{\Phi}(u, v) := \Phi [u, v]_E - [\Phi u, v]_E - [u, \Phi v]_E \, .
\label{derivator}
\end{equation}
In \cite{drinfeldpaper}, we call two calculi each other's \textit{dual} if they satisfy the Jacobi compatibility conditions. For a Lie bialgebroid, the last two of these equations trivially hold, where the first one is equivalent to the compatibility condition (\ref{compatibilityLie}). Hence our resuts from \cite{drinfeldpaper} extend the Lie bialgebroid \cite{mackenzie1994lie} and matched pair of Leibniz algebroid \cite{ibanez2001matched} literatures.

Sometimes algebroids come equipped with certain morphisms to another algebroid structure. For example, $E$-Courant algebroids are equipped with a map to differential operator bundle \cite{Chen_2010}, or the anchor itself can be considered as a map to the tangent Lie algebroid. Given a map
\begin{equation}
    \phi = \phi_A \oplus \phi_Z: E = A \oplus Z \to E' \, ,
\end{equation}
between two algebroids, we see that it is a morphism of brackets in the sense that
\begin{equation}
    [\phi(u), \phi(v)]_{E'} - \phi([u, v]_E) = 0 \, ,
\end{equation}
if the restrictions $\phi_A$ and $\phi_Z$ are morphisims of brackets for $A$ and $Z$ and the following \textit{bracket morphism compatibility condition}
\begin{equation}
    \phi_A(\tilde{K}_{\eta} U) + \phi_Z(\mathcal{L}_U \eta) = [\phi_A(U), \phi_Z(\eta)]_{E'} \, , 
\label{bracketmorphismcompatibilitycondition}
\end{equation}
together with its dual are satisfied \cite{drinfeldpaper}. 

As we have mentioned many algebroids in the literature have a specific form for their bracket's symmetric part \cite{bugden2021g, ccatal2022pre}; namely it can be written in terms of a metric $g_E: E \times E \to \mathbb{E}$ taking values in an arbitrary vector bundle $\mathbb{E}$ and a first-order differential operator $\mathbb{D}_E: \mathbb{E} \to E$:
\begin{equation}
    [u, v]_E + [v, u]_E = \mathbb{D}_E g_E(u, v) \, .
\end{equation}
If the symmetric parts of the brackets on $A$ and $Z$ can be decomposed in this way for some $\mathbb{D}_A, g_A, \mathbb{D}_Z, g_Z$, where the metrics take values in some arbitrary vector bundles $\mathbb{A}$ and $\mathbb{Z}$ then the bracket (\ref{untwistedbracket}) have the symmetric part in terms of the metric
\begin{equation}
    g_E(U + \omega, V + \eta) = g_A(U, V) \oplus \left( \tilde{\iota}_{\omega} V + \tilde{\iota}_{\eta} U \right) \oplus g_Z(\omega, \eta) \oplus \left( \iota_U \eta + \iota_V \omega \right) \, , 
\label{metricdecomp}
\end{equation}
which takes values in the vector bundle 
\begin{equation}
    \mathbb{E} = \mathbb{A} \oplus \mathcal{A} \oplus \mathbb{Z} \oplus \mathcal{Z} \, ,
\end{equation}
and the first-order differential operator $\mathbb{D}_E: \mathbb{E} \to E$
\begin{equation}
    \mathbb{D}_E = \mathbb{D}_A \oplus \tilde{d} \oplus \mathbb{D}_Z \oplus d \, .
\end{equation}
Moreover, just as the anti-symmetry of the Lie bracket forces a relation between right- and left-Leibniz rules, the symmetric part decomposition (\ref{symmetricpart}) forces one to have \cite{ccatal2022pre}
\begin{equation}
    L_E(f, u, v) = \Delta_{\mathbb{D}_E}(f, g_E(u, v)) \, ,
\label{loplop}
\end{equation}
so that the locality operator can be written as \cite{drinfeldpaper}
\begin{equation}
    L_E(f, U + \omega, V + \eta) = L_A(f, U, V) + \Delta_{\tilde{d}} \left( f, \tilde{\iota}_{\omega} V + \tilde{\iota}_{\eta} U \right) \oplus L_Z(f, \omega, \eta) + \Delta_d \left( f, \iota_U \eta + \iota_V \omega \right) \, .
\label{locality}
\end{equation}
Lastly, in order to have metric invariance for the bracket (\ref{untwistedbracket}),
we observe that both $A$ and $Z$ should obey metric invariance property for $g_A$ and $g_Z$:
\begin{equation}
    \mathbb{L}^A_U g_A(V, W) = g_A([U, V]_A, W) + g_A(V, [U, W]_A) \, ,
\end{equation}
and a similar one for $Z$. Moreover the following \textit{metric invariance compatibility conditions} should be satisfied \cite{drinfeldpaper}
\begin{align}
    \pounds_U (\tilde{\iota}_\mu V \oplus \iota_V \mu) &= g_A(V, \tilde{\mathcal{K}}_\mu U) \oplus \tilde{\iota}_\mu [U, V]_A + \tilde{\iota}_{\mathcal{L}_U \mu} V \oplus 0 \oplus \iota_{[U, V]_A} \mu + \iota_V \mathcal{L}_U \mu \, ,  \label{metricinvcond1} \\
    \pounds_U g_Z(\eta, \mu)  &= 0 \oplus \tilde{\iota}_\mu \tilde{\mathcal{K}}_\eta U + \tilde{\iota}_\eta \tilde{\mathcal{K}}_\mu U \oplus g_Z(\mathcal{L}_U \eta, \mu) + g_Z(\eta, \mathcal{L}_U \mu) \oplus \iota_{\tilde{\mathcal{K}}_\eta U} \mu + \iota_{\tilde{\mathcal{K}}_\mu U} \eta \, , \label{metricinvcond2}
\end{align}
together with their duals, where we decompose the metric invariance operator as
\begin{equation}
    \mathbb{L}^E_{U + \omega} = \pounds_U + \tilde{\pounds}_\omega \, .
\end{equation}
We have discussed several algebroid axioms in this section following \cite{drinfeldpaper}, and metric-Bourbaki algebroids \cite{ccatal2022pre} satisfy all of them (except bracket morphism, though its anchor satisfies this property due to the Jacobi identity). Then metric-Bourbaki bialgebroids are defined as a pair of metric-Bourbaki algebroids equipped with calculi satisfying every compatibility condition discussed in this section \cite{drinfeldpaper}. In particular, the Drinfel'd double of a metric-Bourbaki bialgebroid is a metric-Bourbaki algebroid \cite{drinfeldpaper}. This result generalizes the fact that the Drinfel'd double of a Lie bialgebroid is a Courant algebroid \cite{liu1997manin}.

In \cite{ccatal2022pre}, we also defined the notion of Bourbaki calculus with the aim of extending the \v{S}evera classification theorem of exact Courant algebroids to larger class of exact algebroids \cite{vsevera2017letters}. There, we also included the fourth map $\pounds$ in the definition ($\mathcal{L}$ and $\pounds$ are respectively denoted by $\mathcal{L}^Z$ and $\mathcal{L}^R$ in \cite{ccatal2022pre}, while $\iota$ and $d$ are the same). A Bourbaki calculus is then a calculus equipped with a map $\pounds$ further satisfying
\begin{align}
    \pounds_U \iota_V \omega &= \iota_{[U, V]_A} \omega + \iota_V \mathcal{L}_U \omega \, , \nonumber\\
    - \iota_U \mathcal{K}_V \omega &= \iota_V \mathcal{K}_U  \omega \, .
\label{bourbaki}
\end{align}
These properties too are valid for many examples from the literature including Cartan calculus on (almost-)Lie algebroids (see Appendix), and in particular the usual Cartan calculus, so that they are special cases of Bourbaki calculus. The first property is the usual commutation relation between the Lie derivative and the interior product, whereas the second one follows from the Cartan magic formula and the fact that the interior product squares to zero.

%%%%%%%%%%%%%%%%%%%%%%%%%%%%%%%%%%%%%%

\section{$H$- and $R$-Twists}
\label{s5}

This section consists of the main calculations of this study. Now we remove the restriction that both $A$ and $Z$ are subalgebroids of their Drinfel'd double $E = A \oplus Z$, and extend our analysis from the previous section about algebroid axioms. Hence, for the bracket on the vector bundle $E = A \oplus Z$, following the physics nomenclature, we allow $H$- and $R$-twists, which are maps of the form
\begin{equation}
    H: A \times A \to Z \, , \qquad \qquad \qquad R: Z \times Z \to A \, .
\end{equation}
We then examine the effect of the presence of these twists on our calculus framework again for individual algebroid axioms frequently used in the literature. We explicitly derive all the modifications to the definition of calculus and compatibility conditions. We point out the similarities and differences with the untwisted case from the previous section, and explicitly present the ``twisted compatibility conditions''. Our main aim is to pave the path of generalizations of bialgebroids \cite{drinfeldpaper} in the presence of twists and proto Lie bialgebroids \cite{roytenberg2002quasi} to the pairs of vector bundles that are not dual in the usual sense.  

Now, we consider the most general bracket of the form 
\begin{equation}
    [U + \omega, V + \eta]_E = [U, V]_A + \tilde{\mathcal{L}}_{\omega} V + \tilde{\mathcal{K}}_{\eta} U + R(\omega, \eta) \oplus [\omega, \eta]_Z + \mathcal{L}_U \eta + \mathcal{K}_V \omega + H(U, V) \, ,
\label{twistedbracket}
\end{equation}
which we refer to as \textit{Roytenberg bracket}. Similarly to the untwisted case, the anchor is given by
\begin{equation}
    \rho_E = \rho_A \oplus \rho_Z \, .
\end{equation}
In particular, we can decompose this bracket into two ``dual'' parts which we will refer to as twisted Dorfman and twisted tilde-Dorfman brackets:
\begin{align}
    [U + \omega, V + \eta]_{\text{Dor}}^H &= [U, V]_A \oplus \mathcal{L}_U \eta + \mathcal{K}_V \omega + H(U, V) \, ,  \label{twisteddorfmanbracket} \\ 
    [U + \omega, V + \eta]_{\widetilde{\text{Dor}}}^R &= \tilde{\mathcal{L}}_\omega V + \tilde{\mathcal{K}}_\eta U + R(\omega, \eta) \oplus [\omega,\eta]_Z \, . \label{twisteddorfmantildebracket}
\end{align}
Next, we quickly analyze the conditions for the bracket (\ref{twistedbracket}) to satisfy the individual algebroid properties as we did in the previous section, extending our previous results from \cite{drinfeldpaper}. 

The right-Leibniz rule (\ref{rightleibnizrule}) requires both $H$- and $R$-twists are $C^{\infty} M$-linear in the second slot:
\begin{equation}
    H(U, f V) = f H(U, V) \, , \qquad \qquad \qquad R(\omega, f \eta) = f R(\omega, \eta) \, . \label{Hlinearitycond1}
\end{equation}
Similarly, the left-Leibniz rule (\ref{leftleibnizrule}) yields the following conditions for the symbol maps:
\begin{equation}
    \Delta_H^{(1)}(f, U, V) = \text{pr}_Z L_E(f, U, V), \qquad \qquad \qquad \Delta_R^{(1)}(f, \omega, \eta) = \text{pr}_A L_E(f, \omega, \eta) \, . \label{Hlinearitycond2}
\end{equation}
Linearity conditions (\ref{linearityconditions}) for the calculus elements do not change. We also note that in the absence of twists, both (\ref{Hlinearitycond1}) and (\ref{Hlinearitycond2}) are trivially satisfied. We will refer to conditions (\ref{linearityconditions}) together with (\ref{Hlinearitycond1}) and (\ref{Hlinearitycond2}) as \textit{twisted linearity conditions}. Locality operator $L_E$ can be decomposed as
\begin{align}
    L_E(f, U + \omega, V + \eta) &= L_A(f, U, V) + \Delta_{\tilde{\mathcal{K}}}^{(2)}(f, \eta, U) + \rho_Z(\eta)(f) U + \Delta_{\tilde{\mathcal{L}}}^{(1)}(f, \omega, V) + \Delta_R^{(1)}(f, U, \eta) \nonumber\\
    & \quad \oplus L_Z(f, \omega, \eta) + \Delta_{\mathcal{K}}^{(2)}(f, \omega, V) + \rho_A(V)(f) \omega + \Delta_{\mathcal{L}}^{(1)}(f, U, \eta) + \Delta_H^{(1)}(f, U, V) \, .
\label{localitydecomp}
\end{align}

We observe that calculus conditions (\ref{calculusconditions}) coming from the Jacobi identity become:
\begin{align}
    \mathcal{L}_U \mathcal{L}_V \mu - \mathcal{L}_V \mathcal{L}_U \mu - \mathcal{L}_{[U, V]_A} \mu &= -H(U, \tilde{\mathcal{K}}_\mu V) + H(V, \tilde{\mathcal{K}}_\mu U) + [H(U, V), \mu]_Z \, , \nonumber\\
    \mathcal{L}_U d \iota_W \eta - d \iota_{[U, W]_A} \eta - d \iota_W \mathcal{L}_U \eta &= - H(U, \tilde{d} \tilde{\iota}_\eta W) + d_H g_H(\tilde{\mathcal{K}}_\eta U, W) + \mathbb{D}_Z g_Z(H(U, W), \eta) \, , \nonumber\\
    \mathcal{L}_W d \iota_V \omega - d \iota_W d \iota_V \omega &= H(\tilde{d} \tilde{\iota}_\omega V, W) \, .
\label{calculusconditionstwisted}
\end{align}
Together with their duals, these conditions will be referred as \textit{calculus twisted compatibility conditions}. Here we see that the presence of twists imposes stronger conditions so that it is no longer possible to define two separate calculi (in the sense of Section \ref{s4}) anymore. Therefore we name these as compatibility conditions because they mix calculus and tilde-calculus elements together as opposed to the untwisted case. Hence, this indicates that the notion of two individual calculi somewhat dissolves in the presence of twists. Yet, we observe that the conditions (\ref{calculusconditionstwisted}) are only modified through $H$-twist, and $R$-twist does not contribute. Nevertheless as we mentioned earlier, we will still refer to $\mathcal{L}$, $\iota$ and $d$ as calculus elements even when we relax the algebroid axioms further. In the absence of $H$-twist, calculus twisted compatibility conditions (\ref{calculusconditionstwisted}) recover calculus conditions (\ref{calculusconditions}). 

The last two conditions above follow from the $Z$-parts of $(U \eta W)$ and $(\omega V W)$ terms, which have the following bare forms from the Jacobi identity
\begin{align}
    [\mathcal{L}_U, \mathcal{L}_W] \eta - \mathcal{L}_{[U, W]_A} \eta &= \mathcal{L}_U d \iota_W \eta - d \iota_{[U, W]_A} \eta - d \iota_W \mathcal{L}_U \eta \nonumber\\
    & \quad \ - H(\tilde{\mathcal{K}}_{\eta} U, W) - [\eta, H(U, W)]_Z + H(U, \tilde{\mathcal{L}}_{\eta} W) \, , \nonumber\\
    [\mathcal{L}_V, \mathcal{L}_W] \omega  - \mathcal{L}_{[V, W]_A} \omega &= - d \iota_{[V, W]_A} \omega - \mathcal{L}_W d \iota_V \omega - d \iota_W \mathcal{L}_V \omega + d \iota_W d \iota_V \omega + \mathcal{L}_V d \iota_W \omega \nonumber\\
    & \quad \ + H(\tilde{\mathcal{L}}_{\omega} V,  W) + H(V, \tilde{\mathcal{L}}_{\omega} W) - [\omega, H(V, W)]_Z \, . 
\end{align}
The first condition after symmetrization implies
\begin{equation}
    - \mathcal{L}_{\mathbb{D}_A g_A(U, V)} \mu = [d_H g_H(U, V), \mu]_Z \, .
\end{equation}
When $A$ is a Lie algebroid, the left-hand side vanishes due to $g_A = 0$; this resembles the fact that for Courant algebroids $[g_E^{-1} D_E f, u]_E = 0$, which also holds for metric-Bourbaki algebroids \cite{ccatal2022pre} in the form $[\mathbb{D}_E \xi, u]_E = 0$. Moreover, in the absence of $H$-twist, as already observed in \cite{drinfeldpaper}, or for anti-symmetric $H$-twists, the term in the right-hand side vanishes.

In order for the bracket (\ref{twistedbracket}) to satisfy the Jacobi identity (\ref{jacobiidentity}), now we see that the brackets on $A$ and $Z$ should have generally non-vanishing Jacobiators: 
\begin{equation}
    \mathcal{J}_A(U, V, W) = - \tilde{\mathcal{K}}_{H(V, W)} U + \tilde{\mathcal{L}}_{H(U, V)} W + \tilde{\mathcal{K}}_{H(U, W)} V \, . \label{jacobiUVWa}
\end{equation}
Moreover, from the $Z$-part of $(U V W)$ terms, we get the an additional requirement which was absent in the previous section for the twistless case:
\begin{equation}
    H(U, [V, W]_A) - H([U, V]_A, W) - H(V, [U, W]_A) = - \mathcal{L}_U H(V, W) + \mathcal{K}_W H(U, V) + \mathcal{L}_V H(U, W) \, ,
\label{twistjacobih}
\end{equation}
and a similar one holds for the $R$-twist. The Jacobi compatibility conditions (\ref{comp1} - \ref{comp3}) are modified by both $H$- and $R$-twists at the same time:
\begin{align}
    \mathcal{D}^Z_{\mathcal{L}_U}(\eta, \mu) &= \mathcal{L}_{\tilde{\mathcal{K}}_{\eta} U} \mu + \mathcal{K}_{\tilde{\mathcal{K}}_{\mu} U} \eta - H(U, R(\eta, \mu)) \, , \label{comp1twist} \\
    \mathcal{L}_{\tilde{d} \tilde{\iota}_{\eta} U} \mu &= - [d \iota_U \eta, \mu]_Z \, , \label{comp2twist} \\
    d \iota_{\tilde{\mathcal{L}}_{\omega} W} \eta - d \iota_{\tilde{d} \tilde{\iota}_{\eta} W} \omega + d \iota_W [\omega, \eta]_Z &= \mathbb{D}_Z g_Z(d \iota_W \eta, \omega) - \mathbb{D}_Z g_Z(\mathcal{K}_W \omega, \eta) - d_H g_H(R(\omega, \eta), W) \, . \label{comp3twist} 
\end{align}
Here we observe that the second condition remains the same, which is also a crucial property of Dirac structures on Courant algebroids, for instance see Lemma 5.2 in \cite{liu1997manin}. Moreover, we observe that in the absence of $H$-twist, these all become (\ref{comp1} - \ref{comp3}). We will refer to Equations (\ref{jacobiUVWa} - \ref{comp3twist}) together with their duals as \textit{Jacobi twisted compatibility conditions}. 
Equations (\ref{comp1twist} - \ref{comp3twist}) follow from the bare forms of Jacobi identity coming from the $Z$-parts of $(\omega \eta W), (\omega V \mu)$ and $(U \eta \mu)$ terms
\begin{align}
    H(R(\omega, \eta), W) &= [\omega, \mathcal{K}_W \eta]_Z + \mathcal{K}_{\tilde{\mathcal{L}}_{\eta} W} \omega - \mathcal{K}_W ([\omega, \eta]_Z) - [\eta, \mathcal{K}_W \omega]_Z - \mathcal{K}_{\tilde{\mathcal{L}}_{\omega} W} \eta \, , \nonumber\\
    H(V, R(\omega, \mu)) &= [\omega, \mathcal{L}_V \mu]_Z + \mathcal{K}_{\tilde{\mathcal{K}}_{\mu} V} \omega - [\mathcal{K}_V \omega, \mu]_Z - \mathcal{L}_{\tilde{\mathcal{L}}_{\omega} V} \mu - \mathcal{L}_V ([\omega, \mu]_Z) \, , \nonumber\\
    H(U, R(\eta, \mu)) &= - \mathcal{L}_U([\eta, \mu]_Z) + [\mathcal{L}_U \eta, \mu]_Z + \mathcal{L}_{\tilde{\mathcal{K}}_{\eta} U} \mu + [\eta, \mathcal{L}_U \mu]_Z + \mathcal{K}_{\tilde{\mathcal{K}}_{\mu} U} \eta \, ,
\end{align}
respectively.

We continue with the bracket morphism property. For a map $\phi = \phi_A \oplus \phi_Z: E = A \oplus Z \to E'$ to be a bracket morphism, now $\phi_A$ and $\phi_Z$ themselves should not be bracket morphisms, where their failures are measured by the twists:
\begin{equation}
    [\phi_A(U), \phi_A(V)]_{E'} - \phi_A([U, V]_A) = \phi_Z H(U, V) \, , \label{twistedbracketmorphismcomp}
\end{equation}
but the bracket morphism compatibility condition (\ref{bracketmorphismcompatibilitycondition}) remains the same. We will refer to the conditions (\ref{bracketmorphismcompatibilitycondition}) and (\ref{twistedbracketmorphismcomp}) together with their duals as \textit{bracket morphism twisted compatibility conditions}. If $\phi_A$ is a bracket morphism, then this forces $\text{im}(H) \subset \text{ker}(\phi_Z)$.

From the symmetric part, we see that the following terms no longer vanish as in the previous section, when $H$- or $R$-twist are not anti-symmetric:
\begin{equation}
    \text{pr}_Z \mathbb{D}_E g_E(U, V) = H(U, V) + H(V, U) \, , \qquad \quad  \text{pr}_A \mathbb{D}_E g_E(\omega, \eta) = R(\omega, \eta) + R(\eta, \omega) \, , 
\end{equation}
so that the symmetric parts of the twists carry information about the symmetric part of the bracket on $E$, that is $\mathbb{D}_E g_E$. If we enforce that the metric $g_E$ and the map $\mathbb{D}_E$ have similar decompositions as in (\ref{symmetricpart}), we should have some maps
\begin{align}
    g_H: & \ A \times A \to \mathscr{Z} \, , \qquad \qquad \qquad d_H: \mathscr{Z} \to A \, , \nonumber\\
    g_R: & \ Z \times Z \to \mathscr{A} \, , \qquad \qquad \qquad d_R: \mathscr{A} \to Z \, ,
\end{align}
satisfying
\begin{equation}
    H(U, V) + H(V, U) = d_H g_H(U, V) \, , \qquad \quad R(\omega, \eta) + R(\eta, \omega) = d_R g_R(\omega, \eta) \, ,
\label{HRdecomp}
\end{equation}
for some vector bundles $\mathscr{A}$ and $\mathscr{Z}$. In this case the metric $g_E$ decomposes as: 
\begin{equation}
    g_E(U + \omega, V + \eta) = g_A(U, V) \oplus \tilde{\iota}_\omega V + \tilde{\iota}_\eta U \oplus g_R(\omega, \eta) \oplus g_Z(\omega, \eta) \oplus \iota_U \eta + \iota_V \omega \oplus g_H(U, V) \, ,
\label{twistedmetricdecomp}
\end{equation}
which takes values in the vector bundle 
\begin{equation}
    \mathbb{E} = \mathbb{A} \oplus \mathcal{A} \oplus \mathscr{A} \oplus \mathbb{Z} \oplus \mathcal{Z} \oplus \mathscr{Z} \, ,    
\end{equation}
and the first-order differential operator $\mathbb{D}_E: \mathbb{E} \to E$ decomposes as
\begin{equation}
    \mathbb{D}_E = \mathbb{D}_A \oplus \tilde{d} \oplus d_R \oplus \mathbb{D}_Z \oplus d \oplus d_H \,  . \label{twistedfirstorderdecomp}
\end{equation}
Moreover, the symmetric part decomposition (\ref{symmetricpart}) forces one to have \cite{ccatal2022pre}:
\begin{equation}
    L_E(f, u, v) = \Delta_{\mathbb{D}_E}(f, g_E(u, v)) \, .
\end{equation}
This implies the following decomposition for the locality operator
\begin{align}
    L_E(f, U + \omega, V + \eta) &= L_A(f, U, V) + \Delta_{\tilde{d}}(f, \tilde{\iota}_\omega V + \tilde{\iota}_{\eta} U) + \Delta_{d_R}(f, g_R(\omega, \eta)) \nonumber\\
    & \quad \oplus L_Z(f, \omega, \eta) + \Delta_d(f, \iota_U \eta + \iota_V \omega) + \Delta_{d_H}(f, g_H(U, V)) \, .
\label{twistedlocalitydecomp}
\end{align}
This is of course consistent with decomposition (\ref{localitydecomp}) under the assumption of right- and left-Leibniz rules.

Lastly we check the metric invariance property. From $(UVW)$ terms we obtain
\begin{align}
    \pounds_U \left[ g_A(V, W) \oplus g_H(V, W) \right] &= g_A([U, V]_A, W) + g_A(V, [U, W]_A) \oplus \tilde{\iota}_{H(U, V)} W + \tilde{\iota}_{H(U, W)} V \oplus 0 \nonumber\\
    & \quad \oplus 0 \oplus \iota_W H(U, V) + \iota_V H(U, W) \oplus g_H([U, V]_A, W) + g_H(V, [U, W]_A) \, .
\label{twistedmetricinv}
\end{align}
When the $H$-twist vanishes, this is just the metric invariance property (\ref{metricinvariance}) for $A$. Moreover, the metric invariance compatibility conditions (\ref{metricinvcond1}) and (\ref{metricinvcond2}) become 
\begin{align}
    \pounds_U (\tilde{\iota}_\mu V \oplus \iota_V \mu) &= g_A(V, \tilde{\mathcal{K}}_\mu U) \oplus \tilde{\iota}_\mu [U, V]_A + \tilde{\iota}_{\mathcal{L}_U \mu} V \oplus g_R(H(U, V), \mu) \nonumber\\
    & \quad \oplus g_Z(H(U, V), \mu) \oplus \iota_{[U, V]_A} \mu + \iota_V \mathcal{L}_U \mu \oplus g_H(V, \tilde{\mathcal{K}}_\mu U) \, ,  \label{metricinvcond1twist} \\
    \pounds_U \left[ g_R(\eta, \mu) \oplus g_Z(\eta, \mu) \right] &= 0 \oplus \tilde{\iota}_\mu \tilde{\mathcal{K}}_\eta U + \tilde{\iota}_\eta \tilde{\mathcal{K}}_\mu U \oplus g_R(\mathcal{L}_U \eta, \mu) + g_R(\eta, \mathcal{L}_U \mu) \nonumber\\ 
    & \quad \oplus g_Z(\mathcal{L}_U \eta, \mu) + g_Z(\eta, \mathcal{L}_U \mu) \oplus \iota_{\tilde{\mathcal{K}}_\eta U} \mu + \iota_{\tilde{\mathcal{K}}_\mu U} \eta \, . \label{metricinvcond2twist}
\end{align}
We refer to these three conditions (\ref{twistedmetricinv} - \ref{metricinvcond2twist}) together with their duals as the \textit{metric invariance twisted compatibility conditions}. Note that the first and the third ones are modified only by $H$- and $R$-twist respectively, whereas the second one is modified by both twists. We observe that Equations (\ref{metricinvcond1twist}, \ref{metricinvcond2twist}) become (\ref{metricinvcond1}, \ref{metricinvcond2}) when both twists are absent.

We crucially note that twisted compatibility conditions presented in this section recover the versions from the previous section when both $H$- and $R$-twists vanish. Moreover there are some remaining conditions which trivially hold in this case. This will be important in the main proof of Section \ref{s6} where we complete our diagram from the introduction.

Lastly, we remark on one of our previous works \cite{ccatal2022pre}. There, we extended the \v{S}evera classification of exact Courant algebroids to a larger class; namely exact metric-Bourbaki algebroids which fit into the exact sequence \begin{equation}
    0 \xrightarrow{\quad} Z \xrightarrow{\ \quad \chi \ \quad} E \xrightarrow{\ \quad \ \ \quad} A \xrightarrow{\quad} 0 \, ,
\label{exactsequence}
\end{equation}
whose bracket's symmetric part decompose as in Equation (\ref{symmetricpart}) with $\mathbb{D}_E = \chi d$, so that $\mathbb{E} = \mathcal{Z}$. Provided that $A$ is a Lie algebroid and $\text{im}(\chi)$ is $g_E$-isotropic, any exact metric-Bourbaki algebroid have the $H$-twisted Dorfman bracket
\begin{equation} 
    [U + \omega, V + \eta]_{\text{Dor}}^H = [U, V]_A \oplus \mathcal{L}_U \eta - \mathcal{L}_V \omega + d \iota_V \omega + H(U, V) \, ,
\label{ei13}
\end{equation}
and its metric takes the form
\begin{equation}
    g_E(U + \omega, V + \eta) = \iota_U \eta + \iota_V \omega + F(U, V) \, .    
\end{equation}
Then according to Theorem 9.2 of \cite{ccatal2022pre}, the maps $H$ and $F$ satisfy 
\begin{align}
    &H(U, f V) = f H(U, V) \, , \label{ei15a} \\
    &H(f U, V) = f H(U, V) + \Delta_d(f, F(U, V)) \, ,   \label{ei15b} \\
    &H(U, V) + H(V, U) = d F(U, V) \, , \label{ei15c} \\
    &\iota_W H(U, V) + \iota_V H(U, W) = \pounds_U F(V, W) - F([U, V], W) - F(V, [U, W]) \, , \label{ei15d} \\
    &H(U, [V, W]) - H([U, V], W) - H(V, [U, W]) = \mathcal{L}_V H(U, W) - \mathcal{L}_U H(V, W) - \mathcal{L}_W H(U, V) \nonumber\\
    & \qquad \qquad \qquad \qquad \qquad \qquad \qquad \qquad \qquad \qquad + d \iota_W H(U, V) \, . \label{ei15e}
\end{align}
Since $A$ is a Lie algebroid, we have $g_A = 0$, and due to $g_E$-isotropy of $\text{im}(\chi)$, $g_Z = g_R = 0$. So, $\mathbb{D}_A,\mathbb{D}_Z,d_R$ are irrelevant so are their domains $\mathbb{A},\mathbb{Z},\mathscr{A}$. All tilde-calculus elements and $Z$-bracket vanish. Hence $\mathcal{A}$ is irrelevant. After identifying $g_H = F$ and $d_H = d$ so that $\mathscr{Z} \subset \mathcal{Z}$, the third equation (\ref{ei15c}) becomes the decomposition of the symmetric part of $H$. The symbol maps of $H$ are given by the first two equations (\ref{ei15a}, \ref{ei15b}) which are identical to our results in this section by the decompositions of the locality operator given by (\ref{localitydecomp}) and (\ref{twistedlocalitydecomp}). This fact together with the definition of Bourbaki calculus implies the twisted linearity conditions. Condition (\ref{twistedmetricinv}) then yields the fourth condition (\ref{ei15d}), since $g_A = 0$. Condition (\ref{twistjacobih}) is identical to the last condition (\ref{ei15e}), and all other twisted compatibility conditions except for metric invariance are automatically satisfied. For the metric invariance twisted compatibility conditions, the dual of the first condition (\ref{twistedmetricinv}) trivially holds. The second compatibility condition (\ref{metricinvcond1twist}) becomes 
\begin{equation}
    \pounds_U \iota_V \mu = \iota_{[U, V]_A} \mu + \iota_V \mathcal{L}_U \mu \, ,
\end{equation}
which is the one of the defining properties of Bourbaki calculus. Dual of the second one dictates that $\tilde{\pounds}=0$. The third one (\ref{metricinvcond2twist}) is trivially satisfied. Its dual becomes
\begin{equation}
    0 = \iota_W\mathcal{K}_V \omega + \iota_V \mathcal{L}_W \omega
\end{equation}
which is the other defining property of Bourbaki calculus. In conclusion, we see that our results from this section recover Theorem 9.2 of \cite{ccatal2022pre} about exact metric-Bourbaki algebroids, which we presented above. When isotropic splittings exist as in the case of Courant, conformal Courant \cite{baraglia2013conformal} and $AV$-Courant \cite{libland2011} algebroids, one can choose $F = 0$ so that $H$ is $C^{\infty}M$-bilinear and anti-symmetric. Consequently, $H$ is a $Z$-valued 2-form on $A$ as observed in Corollary 9.1 of \cite{ccatal2022pre}. Moreover for Courant algebroids, the condition (\ref{ei15d}) from metric invariance in this case implies that $H$ is indeed a usual 3-form, and the last condition (\ref{ei15e}) from the Jacobi identity implies that~$H$ is closed with respect to the usual exterior derivative.

%%%%%%%%%%%%%%%%%%%%%%%%%%%

\section{Proto Bialgebroids}
\label{s6}

In the prelude, Section \ref{s3}, we discussed Lie bialgebroids and their twistful generalizations proto Lie bialgebroids. This relation corresponds to the lower arrow in the diagram that we discussed in the introduction. Both of these structures are defined on a pair of vector bundles which are dual to each other; and their Drinfel'd double yields a Courant algebroid structure. With the motivation that U-duality extensions of ideas coming from T-duality necessitate the use of pairs which are not necessarily dual and of arbitrary ranks, we extended Lie bialgebroids to bialgebroids in Section \ref{s4}. This extension corresponds to the left arrow of our diagram. In this section, which includes the main results of the paper, our aim is two-fold: On the one hand we generalize the notion of bialgebroids that we defined in our earlier work \cite{drinfeldpaper} to include both $H$- and $R$-twists. On the other hand we generalize the proto Lie bialgebroids \cite{roytenberg2002quasi} to a pair of algebroids which are not dual. Here, in light of the analysis on algebroid axioms in Section \ref{s5}, we collect our observations and introduce the notion of proto bialgebroids for a pair of vector bundles that are not necessarily dual. This generalization corresponds to the upper arrow in the diagram. The remaining right arrow is the one that is most involved. To this end, we explicitly show that proto Lie bialgebroids of Roytenberg \cite{roytenberg2002quasi} satisfy all of the twisted compatibility conditions that we derived in the previous section, so that we prove they can be considered as proto metric-Bourbaki bialgebroids which we also introduce in this section extending our earlier works \cite{ccatal2022pre, drinfeldpaper}. We present our results without referring to the supermanifold formalism as usually done in the literature, in a ``bosonic'' language in the spirit of \cite{chatzistavrakidis2015sigma} on which we also comment in details. We also point out important special cases where only one of the twists is present, which generalizes the quasi Lie and Lie quasi bialgebroids of \cite{roytenberg2002quasi}. The notion of proto bialgebroid is expected to be relevant for the exceptional field theories \cite{pacheco2008m, hohm2013exceptional, Berman:2020tqn}, higher dimensional exceptional Drinfel'd algebras \cite{sakatani2020u, malek2020poisson, blair2022generalised, malek2021e6, blair2020exploring} and extensions of (Nambu-Lie) U-duality \cite{sakatani2020non, sakatani2021extended, musaev2022non} since all of these depend on the exceptional geometries on vector bundles which are in the form of a direct sum of two or more non-dual vector bundles like the ones in Equation (\ref{physicsbundles}). Now we introduce proto bialgebroids which can be considered as U-duality extensions of proto Lie bialgebroids of \cite{roytenberg2002quasi} and twistful generalizations of our bialgebroids \cite{drinfeldpaper}. With the notion of proto bialgebroid, we now complete our diagram:

\[\begin{tikzcd}
	{\quad \mbox{Bialgebroids of \cite{drinfeldpaper}}\quad } &&&&&& {\quad \mbox{Proto bialgebroids} \quad} \\
	\\
	\\
	{\mbox{\ Lie bialgebroids of \cite{mackenzie1994lie}} \ } &&&&&& {\ \mbox{Proto Lie bialgebroids of \cite{roytenberg2002quasi}} \, . \ }
	\arrow["{\mbox{Twisted comp. conditions}}", from=1-1, to=1-7]
	\arrow["{\mbox{$A^*$ to $Z$}}", from=4-1, to=1-1]
	\arrow["{\ \ \ \ \ \ \ \ \ \ \ \ \mbox{Twisted comp. conditions} \ \ \ \ \ \ \ \ }", from=4-1, to=4-7]
	\arrow["{\mbox{$A^*$ to $Z$}}"', from=4-7, to=1-7]
\end{tikzcd}\]

In the previous section, we have derived the necessary compatibility conditions for each algebroid property in the presence of both twists. As these axioms are analyzed individually, we may define the notion of proto bialgebroid in general. We term the pair $(A, Z)$ a \textit{proto bialgebroid} if both $A$ and $Z$ are algebroids equipped with calculus elements and twists satisfying a desired set of twisted linearity and twisted compatibility conditions together with their duals. For a proto bialgebroid $(A, Z)$, we call the algebroid structure on $E = A \oplus Z$ given by the bracket (\ref{twistedbracket}) as the proto bialgebroid's \textit{Drinfel'd double}. Then our main result from the previous section can be summarized as the Drinfel'd double of a proto bialgebroid satisfying a set of twisted linearity and compatibility conditions itself satisfies the related algebroid properties. Since we included the duals of every twisted linearity and compatibility condition, we automatically have $(Z, A)$ is also a proto bialgebroid. In particular, one may define the notion of \textit{proto metric-Bourbaki bialgebroid} as a proto bialgebroid whose twists can be decomposed as in Equation (\ref{HRdecomp}) which satisfy every twisted linearity and compatibility condition we determined and stated. Then we also have that the Drinfel'd double of a proto metric-Bourbaki bialgebroid is a metric-Bourbaki algebroid. In Section \ref{s5}, we observed that turning the $H$- and $R$-twists off by setting $H = R = 0$, every twisted linearity and compatibility condition recovers their twistless versions from Section \ref{s4} in addition to some conditions that trivially hold. This in particular implies that in the absence of twists, the notion of proto bialgebroids reduces to the notion of bialgebroids that we introduced \cite{drinfeldpaper}. This effectively explains both of the horizontal arrows of the diagram. As we have explicitly proved in \cite{drinfeldpaper}, when we choose $Z = A^*$, the notion of metric-Bourbaki bialgebroid reduces to the usual notion of Lie bialgebroids, which explains the left arrow.

Now we prove that proto Lie bialgebroids of \cite{roytenberg2002quasi} are special cases of proto metric-Bourbaki bialgebroids, which completes the right arrow of our diagram. The proto Lie bialgebroids that we presented in Section \ref{s3} using the supermanifold formalism can be also formulated in a ``bosonic'' language as in \cite{chatzistavrakidis2015sigma}. In order to relate the notation of the work \cite{roytenberg2002quasi} to ours, we make the identifications
\begin{equation}
    A = A \, , \quad Z = A^* \, , \quad \rho_A = a \, , \quad \rho_{A^*} = a_* \, , \quad H = \phi \, , \quad R = \phi \, , \quad d = d_{\mu} \, , \quad \tilde{d} = d_{\gamma} \, ,
\end{equation}
and the calculus elements are constructed from the brackets $[\cdot,\cdot]_A$ and $[\cdot,\cdot]_{A^*}$ as in Appendix. We also make a comparison with the results of \cite{chatzistavrakidis2015sigma}, and the relation to our notation is given as
\begin{equation}
    A = L \, , \quad Z = L^* \, , \quad \rho_A = \rho \, , \quad \rho_{A^*} = \rho_* \, , \quad H = \phi \, , \quad R = \psi \, , \quad d = d_L \, , \quad \tilde{d} = d_{L^*} \, .
\end{equation}
We also note that there is a convention difference with a minus sign when the multivectors are fed into the forms.

From now on we will use $A^*$ for the algebroid $Z$ because it is dual to $A$. Moreover $\mathcal{A}=\mathcal{Z}=C^\infty M$ since $\iota$ and $\tilde{\iota}$ are both canonical pairings.  The effects of the twists can be realized as the modification of the usual properties of algebroids $A$ and $A^*$, which are not Lie algebroids anymore. For example, for a proto Lie bialgebroid, the anchors are not morphisms of brackets as given in Equation (3.3) of \cite{roytenberg2002quasi} and Property (2) in Definition 2.1 of \cite{chatzistavrakidis2015sigma}:
\begin{equation}
    \rho_A([U, V]_A) = [\rho_A(U), \rho_A(V)]_{\text{Lie}} + \rho_{A^*}(H(U, V)) \, ,
\end{equation}
which corresponds to our Equation (\ref{twistedbracketmorphismcomp}) for $\phi_A = \rho_A, \phi_Z = \rho_{A^*}$. We will not present the conditions for $Z$ because they are completely analogous to the ones for $A$; they are just duals. Moreover, the Jacobi identities are twisted as in Equation (3.4) of \cite{roytenberg2002quasi} and Property (3) in Definition 2.1 of \cite{chatzistavrakidis2015sigma} (with a plus sign in the third term due to the sign convention difference) which is equivalent to
\begin{equation}
    \mathcal{J}_A(U, V, W) = \tilde{d}(H(U, V, W)) + H(\tilde{d} U, V, W) - H(U, \tilde{d} V, W) + H(U, V, \tilde{d} W) \, ,
\label{minusproblem}
\end{equation}
 due to anti-symmetry of the bracket on $A$. Here, the $H$-twist is also considered as a bundle map $\Lambda^4 A \to A$. This equation corresponds to our Equation (\ref{jacobiUVWa}) by using the definitions of $\mathcal{L}$ and $d$ from the Appendix and
\begin{equation}
    H(\tilde{d}U, V, W)(\mu) = \tilde{d}U(H(V, W), \mu) = - \tilde{\iota}_{\mu} \tilde{\iota}_{H(V, W)} \tilde{d} U = (\mathcal{K}_{H(V, W)} U)(\mu) \, .
\end{equation}
Moreover, the algebroids satisfy the right-Leibniz rule, as noted in Property (1) in Definition 2.1 of \cite{chatzistavrakidis2015sigma}, so that both $A$ and $A^*$ are almost-Lie algebroids since they have anti-symmetric brackets. Due to this anti-symmetry, we have $g_A = g_{A^*} = 0$ so that $\mathbb{D}_{A}$, $\mathbb{D}_{A^*}$, $\mathbb{A}$ and $\mathbb{Z}$ (which denotes the range of the metric on $A^*$) are irrelevant. Similarly, both $H$ and $R$ are totally anti-symmetric so that we have $g_H = g_R = 0$ and $d_H, d_R, \mathscr{Z}$ (which is associated to the symmetric part decomposition of $H$) and $\mathscr{A}$ are irrelevant. On the other hand, the $H$-twist should be closed in the sense that $d H = 0$ as in Property (4) in Definition 2.1 of \cite{chatzistavrakidis2015sigma}, which is equivalent to Equation (3.5) of \cite{roytenberg2002quasi}
\begin{align}
    &[H(U, V, W), Y]_{\text{SN},A} - [H(U, V, Y), W]_{\text{SN},A}  + [H(U, W, Y), V]_{\text{SN},A} - [H(V, W, Y), U]_{\text{SN},A}  \nonumber\\ 
    &\qquad - H([U, V]_A, W, Y) + H([U, W]_A, V, Y) - H([U, Y]_A, V, W) \nonumber\\
    &\qquad - H([V, W]_A, U, Y) + H([V, Y]_A, U, W) - H([W, Y]_A, U, V) = 0 \, .
\end{align}
This corresponds to our Equation (\ref{twistjacobih}) by first noting $[H(U, V, W), Y]_{\text{SN},A}=-\rho_A(Y)(H(U,V,W))$ and using the definitions presented in Appendix. The fundamental compatibility condition $\{ \mu, \gamma \}=0$, which reduces to the usual compatibility condition (\ref{compatibilityLie}) for Lie bialgebroids \cite{mackenzie1994lie} is then equivalent to our Equation (\ref{comp3twist}). These together with their duals exhaust all the conditions presented in Equations (\ref{roytenberg1} - \ref{roytenberg5}) \cite{roytenberg2002quasi}. Yet, we have several extra twisted compatibility conditions remaining, so that we need to prove all of these extra conditions are satisfied for proto Lie bialgebroids.  

We first note that twisted linearity conditions and their duals all hold due to Cartan magic formula and the twists being tensorial. As noted above, $g_A = g_{A^*} = g_H = g_R = 0$. This simplifies several conditions. For example, the second calculus twisted compatibility condition in (\ref{calculusconditionstwisted}) becomes
\begin{equation}
    \mathcal{L}_U d \iota_W \eta - d \iota_{[U, W]_A} \eta - d \iota_W \mathcal{L}_U \eta = - H(U, \tilde{d} \tilde{\iota}_\eta W) = (\tilde{d} \tilde{\iota}_{\eta} W)(H(U, V)) = \rho_Z(H(U, V))(\iota_W \eta) \, ,
\end{equation}
where the left-hand side paired with $V \in A$ explicitly reads by the first equation in (\ref{calculuslie})
\begin{equation}
    (\mathcal{L}_U d \iota_W \eta)(V) - (d \iota_{[U, W]_A} \eta)(V) - (d \iota_W \mathcal{L}_U \eta)(V) = \mathcal{P}_{\rho_A}(U, V)(\iota_W \eta) \, ,
\end{equation}
where we define the \textit{predator} of the anchor $\rho_E$ as a map $\mathcal{P}_{\rho_E}: E \times E \to TM$ given by
\begin{equation}
    \mathcal{P}_{\rho_E}(u, v) := [\rho_E(u), \rho_E(v)]_{\text{Lie}} - \rho_E([u, v]_E) \, .
\end{equation}
For a proto Lie bialgebroid, one has 
\begin{equation}
    \mathcal{P}_{\rho_A}(U, V) = \rho_Z H(U, V) \, ,
\label{predatorofA}
\end{equation}
so the second calculus twisted compatibility condition in (\ref{calculusconditionstwisted}) holds. Similarly, the third one in (\ref{calculusconditionstwisted}) follows by the second equation in (\ref{calculuslie}). For the first condition in (\ref{calculusconditionstwisted}), we observe that by the first equation in (\ref{calculuslie-oneform}) we have
\begin{equation}
    \mathcal{L}_U \mathcal{L}_V \mu (W) - \mathcal{L}_V \mathcal{L}_U \mu (W) - \mathcal{L}_{[U, V]_A} \mu (W) = \mathcal{P}_{\rho_A}(U, V) (\iota_W \mu) - \mu (\mathcal{J}_A(U, V, W)) \, .
\end{equation}
By using the Jacobiator of $A$ given by (\ref{jacobiUVWa}), the predator given by (\ref{predatorofA}), and the definitions of $\tilde{\mathcal{L}}$ and $\tilde{d}$ from the Appendix, we can see that this condition is also satisfied. The duals of these conditions are also valid for the tilde-calculus elements since $A^*$ is also an almost-Lie algebroid.

The first Jacobi twisted compatibility condition (\ref{comp1twist}) together with its dual are coming from the compatibility condition of proto Lie bialgebroids given by (\ref{roytenberg3}). The second Jacobi twisted compatibility condition (\ref{comp2twist}) is implied by the first one (\ref{comp1twist}), and its dual follows similarly. To see this one replaces $U$ by $f U$ in the first condition (\ref{comp1twist}) and use the twisted linearity conditions and their duals for the calculus elements and the Leibniz rule for the bracket. The third Jacobi twisted compatibility condition (\ref{comp3twist}) together with its dual directly follow from the definitions of calculus elements outlined in the Appendix. All metric invariance twisted compatibility conditions (\ref{twistedmetricinv} - \ref{metricinvcond2twist}) together with their duals hold because $g_A = g_{A^*} = g_H = g_R = 0$ and the usual commutation relation between $\iota, \mathcal{L}$ and $\tilde{\iota}, \tilde{\mathcal{L}}$ hold as in the fourth equation of (\ref{cartancalculusrelations}) after identifying $\pounds=\mathcal{L}$ and $\tilde{\pounds}=\tilde{\mathcal{L}}$. This follows from the fact that two interior products coincide, \textit{i.e.}, $\iota_V \omega = \tilde{\iota}_{\omega} V$, and square to zero since they are just the canonical pairing, and both $H$ and $R$ are totally anti-symmetric so that $\iota_W H(U, V) + \iota_V H(U, W) = 0$. As we have explicitly showed in the above discussion every twisted linearity and twisted compatibility conditions that we have derived in Section \ref{s5} hold for proto Lie bialgebroids. 

Finally, there is one last condition that we need to discuss which completes the proof that every proto Lie bialgebroid is a proto metric-Bourbaki bialgebroid. This compatibility condition seems to be missing in the ``bosonic'' version of proto Lie bialgebroids in Definition 2.1 of \cite{chatzistavrakidis2015sigma} based on their earlier work \cite{chatzistavrakidis2014dirac}. This definition does not include the first Jacobi twisted compatibility condition (\ref{comp1twist}), which reduces to the usual compatibility condition (\ref{compatibilityLie}). Even though in both \cite{roytenberg2002quasi} and \cite{roytenberg1999courant} the condition $\{ \mu, \gamma \} = 0$ yields  that the exterior derivative induced by the bracket on $A^*$ is a derivation of the Schouten-Nijenhuis bracket induced by the bracket on $A$, this is also missing in the bosonic version of the conditions defining quasi Lie bialgebroids \cite{roytenberg1999courant, roytenberg2002quasi}. Nevertheless, we note that this condition is included in the definition of quasi Jacobi bialgebroids \cite{da2006twisted} in a graded setting. Although this condition is missing in \cite{chatzistavrakidis2015sigma}, it is noted that one can directly check that the Drinfel'd double still yields a Courant algebroid structure for their explicit example. We can indeed check that the first Jacobi twisted compatibility condition holds for the example constructed in \cite{chatzistavrakidis2015sigma}. For this example, we consider two almost-Lie algebroids $(A, \rho_A, [\cdot,\cdot]_A)$ and $(A^*, \rho_{A^*}, [\cdot,\cdot]_{A^*})$ together with tensors $H \in \Lambda^3 A^*$, $\Pi \in \Lambda^2 TM$, and $B \in \Lambda^2 T^*M$. The vector bundles $A$ and $A^*$ are defined by $\Psi^{-1} TM$ and $\Psi^{-1} T^*M$ respectively, where $\Psi = e^{-\Pi} e^{-B}$ is considered as maps of the form $A \to TM$ and $A^* \to T^*M$. The anchors are given by $\rho_A = \Psi$ and $\rho_{A^*} = \Pi \Psi$. In order to have a proto Lie bialgebroid structure, the following brackets are introduced
\begin{align}
    [U, V]_A &= \Psi^{-1}[\Psi U, \Psi V]_{\text{Lie}} + \Psi^{-1} \Pi \Psi H(U, V) \\
    [\eta, \mu]_{A^*} &= \Psi^{-1}[\Psi \eta, \Psi \mu]_{\text{Kos}} + H(\Psi^{-1} \Pi \Psi \eta, \Psi^{-1} \Pi \Psi \mu) \, ,
\end{align}
where the Koszul bracket in terms of the bivector $\Pi$ is defined on $T^*M$ by \cite{koszul1985crochet, bi2011higher}
\begin{equation}
    [\alpha, \beta]_{\text{Kos}} := \mathcal{L}_{\Pi \alpha} \beta - \mathcal{L}_{\Pi \beta} \alpha + d \iota_{\Pi \beta} \alpha \, , \label{koszulbracket}
\end{equation}
for $\alpha, \beta \in T^*M$ in terms of the Cartan calculus elements $(\mathcal{L}, \iota, d)$ constructed from the bracket on $A$. Finally, the $R$-twist is chosen as an element $R \in \Lambda^3 A$ of the form
\begin{equation}
    R(\eta, \mu) = \tfrac{1}{2} \Psi^{-1}[\Pi, \Pi]_{\text{SN,Lie}}(\Psi \eta, \Psi \mu) \, ,
\end{equation}
where $[\cdot,\cdot]_{\text{SN,Lie}}$ is the Schouten-Nijenhuis bracket constructed from the Lie bracket on $TM$. With this setup, we are able to prove on a frame that the first Jacobi twisted compatibility condition (\ref{comp1twist}) holds. In order to see this, one should express everything on a frame and use $C^{\infty}M$-linearity properties for the above maps. Moreover, one needs to use $d H = 0$,  which is included in the definition, and $\Pi^3 H = 0$, in the sense that $H(\Pi \omega, \Pi \eta, \Pi \mu) = 0$ which is assumed in \cite{chatzistavrakidis2015sigma}. Lastly, the nilmanifold assumption of \cite{chatzistavrakidis2015sigma} clears all the extra terms between two equations, and one gets the desired result. This means that the structure functions of the tangent Lie algebroid are two-step nilpotent, \textit{i.e.}, $f_{a b}{}^c f_{c d}{}^e = 0$ with no sum on the repeated index. Hence with two additional requirements about niloptency and $\Pi^3 H = 0$ which are assumed in \cite{chatzistavrakidis2015sigma}, their specific example is a proto Lie bialgebroid whose Drinfel'd double is a Courant algebroid. Finally, we would like to note that, since all the twisted linearity and compatibility conditions are satisfied, it is an example of proto metric-Bourbaki bialgebroid whose Drinfel'd double is a metric-Bourbaki algebroid. 

Next, we can consider special subcases of proto bialgebroids where one of the twists vanishes. These are analogous to the quasi Lie and Lie quasi bialgebroids of \cite{roytenberg2002quasi}. We call proto bialgebroids with $R = 0$ as quasi bialgebroids, and the ones with $H = 0$ as tilde-quasi bialgebroids, which extend quasi Lie and Lie quasi bialgebroids, respectively. In particular, for the case of proto metric-Bourbaki bialgebroids, we would have quasi metric-Bourbaki and tilde-quasi metric-Bourbaki bialgebroids, whose specific examples will be noted in Section \ref{s8}. For quasi and tilde-quasi bialgebroids, some of the conditions simplify; we will only comment on the latter since the former is completely analogous. Since $H = 0$ for a tilde-quasi bialgebroid, twisted linearity conditions related to the $H$-twist trivially hold. Moreover, the calculus twisted compatibility conditions (\ref{calculusconditionstwisted}) coincide with the untwisted versions (\ref{calculusconditions}), so that the triplet $(\mathcal{L}, \iota, d)$ defines a calculus in the sense of Section \ref{s4}. Also the bracket on $A$ satisfies the Jacobi identity as Equation (\ref{jacobiUVWa}) does not get modified. Consequently, $\rho_A$ is a bracket morphism as can be also noted from Equation (\ref{twistedbracketmorphismcomp}) after identifying $\phi_A$ with $\rho_A$. Equation (\ref{twistjacobih}) trivially gets satisfied. Furthermore, twisted Jacobi compatibility conditions (\ref{comp1twist} - \ref{comp3twist}) together with their duals do not get modified so they just reduce to the usual Jacobi compatibility conditions. The terms containing $H$, and hence $g_H$, disappear from the decompositions of metric (\ref{twistedmetricdecomp}), the locality operator (\ref{twistedlocalitydecomp}), and twisted metric compatibility conditions. The twisted metric invariance compatibility condition for $A$ (\ref{twistedmetricinv}), together with the condition (\ref{metricinvcond1twist}) become the usual metric invariance property for $A$, and the usual metric invariance compatibility condition (\ref{metricinvcond1}), respectively. For quasi bialgebroids, due to the symmetry between $A$ and $Z$, the ``duals'' of above comments hold true since in this case $R = 0$.  It is interesting to note that, for both quasi  and tilde-quasi bialgebroids, Jacobi twisted compatibility conditions (\ref{comp3twist}) are the same and they both reduce to the usual Jacobi compatibility conditions (\ref{comp1} - \ref{comp3}). 

An important class of examples contain twists which are anti-symmetric and $C^{\infty}M$-bilinear as in the case of proto Lie bialgebroids. In this case $g_H = g_R = 0$, so that $d_H, d_R, \mathscr{Z}$ and $\mathscr{A}$ are irrelevant and some simplifications occur.  We observe that when the twists are anti-symmetric, the decompositions of the locality operator (\ref{twistedlocalitydecomp}) and metric (\ref{twistedmetricdecomp}) are the same as the untwisted results (\ref{locality}) and (\ref{metricdecomp}), respectively. Most notably, the third Jacobi twisted compatibility condition (\ref{comp3twist}) reduces to the usual one (\ref{comp3}). Moreover metric invariance twisted compatibility condition (\ref{twistedmetricinv}) reduces to 
\begin{align}
    \pounds_U g_A(V, W) &= g_A([U, V]_A, W) + g_A(V, [U, W]_A) \oplus \tilde{\iota}_{H(U, V)} W + \tilde{\iota}_{H(U, W)} V \oplus 0 \nonumber\\
    & \quad \oplus 0 \oplus \iota_W H(U, V) + \iota_V H(U, W) \oplus 0 \, ,
\label{twistedmetricinvantisymm}
\end{align}
which implies that the $H$-twist somewhat measures the deviation of metric invariance of $g_A$. Also, the dual of (\ref{metricinvcond2twist}) does not get modified, and reduce to the usual one (\ref{metricinvcond2}).

%%%%%%%%%%%%%%%%%%%%%%%%%%%%%%%%%%%%%%%%

\section{Brackets Induced by Twist Automorphisms}
\label{s7}

\noindent The main interest of this section is the transformation of brackets where we consider a general twist automorphism, which usually appears in the physics nomenclature in the form of a twist matrix on a suitably chosen frame. We perform our calculations in the most general setting between two Roytenberg brackets in a frame independent manner. In particular, we present the transformation rules for calculi and twists induced by a vector bundle automorphism. We then focus on three special cases of automorphisms which extend frame changes, $B$-field and $\beta$-transformations.

Such twist automorphisms can be used to generate missing terms in a bracket of the form~\ref{twistedbracket}. In most cases, one usually has only the (twisted) Dorfman bracket expressed in terms of usual Cartan calculus. A natural question is whether the remaining terms can be constructed using automorphisms. For example, a $B$-transformation may be used to induce an $H$-twist to the Dorfman bracket with $H = dB$. When $B$ is closed, then the Dorfman bracket is preserved, and in general this is not true. For instance a $\beta$-transformation introduces additional terms to the Dorfman bracket. In this case the Dorfman bracket is not preserved. However, when the Q flux associated with the $\beta$-transformation vanishes, the bracket will remain the same for certain 'invariant' sections. This is exactly what happens in Yang-Baxter deformations, which will be discussed in an upcoming paper. In fact, using such a $\beta$-transformation one can generate a new supergravity solution starting from a given one in the DFT framework \cite{Sakamoto:2017cpu, Sakamoto:2018krs, Catal-Ozer:2019tmm}. Moreover, $\beta$-transformations are intimately related to non-geometric backgrounds in DFT \cite{Fernandez-Melgarejo:2017oyu, asakawa2015poisson, andriot2013beta}. The effect of twist matrices that depend on both $B$- and $\beta$-transformations are considered in \cite{davidovic2021courant}. Furthermore, fluxes associated with most general twist matrices have been studied in \cite{chatzistavrakidis2018double, tomas2020generalized, coimbra2011supergravity}. 

Twist matrices are used to construct certain sections $(E_A)$ of the generalized tangent bundle, whose ``generalized Lie derivatives'' via
\begin{equation}
    \hat{\mathcal{L}}_{E_A} E_B = F_{AB}{}^C E_C\, , \label{generalizedlie}
\end{equation}
yield the fluxes $F_{AB}{}^C$ where $\hat{\mathcal{L}}$ denotes the generalized Lie derivative induced by the Dorfman bracket. These generalized vielbeins, and their exceptional counterparts, are a key ingredient in generalized Scherk-Schwartz reductions of DFT and ExFT that encode the effects of fluctuations on the background \cite{aldazabal2011effective, dabholkar2003duality, Catal-Ozer:2017cqr, Catal-Ozer:2017ycb}. Particularly, they encapsulate the dependence of generalized tensors to coordinates on the internal space. These generalized vielbeins can be used to construct a generalized metric, which can be parametrized in terms of the bosonic field content of the low energy effective theory. For instance in \cite{halmagyi2009non}, different parametrizations of generalized metric are studied using $B$- and $\beta$-transformations extensively based on their earlier work on current algebras \cite{halmagyi2008non}. A Roytenberg bracket is constructed from an initial twisted Dorfman bracket using a $\beta$-transformation for a bivector $\beta$ via:
\begin{equation}
    [U + \omega, V + \eta]_{\text{Roy}} = e^{-\beta} [e^\beta (U + \omega), e^\beta(V + \eta)]^H_{\text{Dor}} \, . 
\label{halmagyitrick}
\end{equation} 
This procedure, on a frame, agrees with the generalized Lie derivatives of generalized vielbeins as in (\ref{generalizedlie}) realizing the fluxes $F_{A B}{}^C$ as the structure constants of the Roytenberg bracket. In \cite{drinfeldpaper}, we extended this procedure, which can be considered as a generalization of triangular Lie bialgebroids, to the realm of higher Courant algebroids \cite{bi2011higher} and Nambu-Poisson structures \cite{nambu1973generalized, takhtajan1994foundation}.

Now, we present the most general form of such a procedure (\ref{halmagyitrick}) between two Roytenberg brackets
\begin{equation}
    [u, v]_{\text{Roy}'} = \Psi^{-1} [\Psi u, \Psi v]_{\text{Roy}}\, ,
\label{brackettrick}
\end{equation}
induced by a twist automorphism 
$\Psi: E = A \oplus Z \to E = A \oplus Z$. Here, the initial bracket $[\cdot,\cdot]_{\text{Roy}}$ have all 8 components as in Equation (\ref{twistedbracket}) and the twist $\Psi\in \text{Aut}(E)$ is of the general form which can be decomposed as
\begin{equation}
    \Psi = \begin{pmatrix}
                \psi_{11} & \psi_{12} \\
                \psi_{21} & \psi_{22}
            \end{pmatrix}\, ,
\label{psiform}
\end{equation}
where $\psi_{11} \in \text{Aut}(A)$, $\psi_{22} \in \text{Aut}(Z)$, $\psi_{12}: Z \to A$ and $\psi_{21}: A \to Z$. The action of $\Psi$ on a section $U + \omega \in E = A \oplus Z$ takes the form:
\begin{equation}
\Psi(U + \omega) = 
\begin{pmatrix}
\psi_{11} & \psi_{12} \\
\psi_{21} & \psi_{22}
\end{pmatrix}
\begin{pmatrix}
U \\
\omega  
\end{pmatrix} 
=
\begin{pmatrix}
\psi_{11} U + \psi_{12} \omega \\
\psi_{21} U + \psi_{22} \omega  
\end{pmatrix} 
= \psi_{11} U + \psi_{12} \omega \oplus \psi_{21} U + \psi_{22} \omega \, . 
\end{equation}

\noindent If we denote the inverse of the automorphism $\Psi$ by
\begin{equation}
	\Psi^{-1} = \begin{pmatrix}
					\phi_{11} & \phi_{12} \\
					\phi_{21} & \phi_{22}
				\end{pmatrix} \, ,
\end{equation}
the method (\ref{brackettrick}) induces modifications of the components of the initial bracket as follows. Extending the simple form of Equation (\ref{brackettrick}) for different components, we get the following lengthy expressions. The brackets on $A$ and $Z$ are modified as
\begin{align}
    [U, V]'_A &= \phi_{11} \left\{ [\psi_{11} U, \psi_{11} V]_A + \tilde{\mathcal{L}}_{\psi_{21} U} \psi_{11} V + \tilde{\mathcal{K}}_{\psi_{21} V} \psi_{11} U + R(\psi_{21} U, \psi_{21} V) \right\} \nonumber\\
    & \quad \ + \phi_{12} \left\{ [\psi_{21} U, \psi_{21} V]_Z + \mathcal{L}_{\psi_{11} U} \psi_{21} V + \mathcal{K}_{\psi_{11} V} \psi_{21} U + H(\psi_{11} U, \psi_{11} V) \right\} \, , \nonumber\\
    [\omega,\eta]'_Z &= \phi_{21} \left\{ [\psi_{12} \omega, \psi_{12} \eta]_A + \tilde{\mathcal{L}}_{\psi_{22} \omega} \psi_{12} \eta + \tilde{\mathcal{K}}_{\psi_{22} \eta} \psi_{12} \omega + R(\psi_{22} \omega, \psi_{22} \eta) \right\} \nonumber\\
    & \quad \ + \phi_{22} \left\{ [\psi_{22} \omega, \psi_{22} \eta]_Z + \mathcal{L}_{\psi_{12} \omega} \psi_{22} \eta + \mathcal{K}_{\psi_{12} \eta} \psi_{22} \omega + H(\psi_{12} \omega, \psi_{12} \eta) \right\} \, ,
\end{align}
and the modifications of $H$- and $R$-twists read
\begin{align}
    H'(U, V) &= \phi_{21} \left\{ [\psi_{11} U, \psi_{11} V]_A + \tilde{\mathcal{L}}_{\psi_{21} U} \psi_{11} V + \tilde{\mathcal{K}}_{\psi_{21} V} \psi_{11} U + R(\psi_{21} U, \psi_{21} V) \right\} \nonumber\\
    & \quad \ + \phi_{22} \left\{ [\psi_{21} U, \psi_{21} V]_Z + \mathcal{L}_{\psi_{11} U} \psi_{21} V + \mathcal{K}_{\psi_{11} V} \psi_{21} U + H(\psi_{11} U, \psi_{11} V) \right\} \, , \nonumber\\
    R'(\omega, \eta) &= \phi_{11} \left\{ [\psi_{12} \omega, \psi_{12} \eta]_A + \tilde{\mathcal{L}}_{\psi_{22} \omega} \psi_{12} \eta + \tilde{\mathcal{K}}_{\psi_{22} \eta} \psi_{12} \omega + R(\psi_{22} \omega, \psi_{22} \eta) \right\} \nonumber\\
    & \quad \ + \phi_{12} \left\{ [\psi_{22} \omega, \psi_{22} \eta]_Z + \mathcal{L}_{\psi_{12} \omega} \psi_{22} \eta + \mathcal{K}_{\psi_{12} \eta} \psi_{22} \omega + H(\psi_{12} \omega, \psi_{12} \eta) \right\} \, .
\end{align}
On the other hand, we observe that the calculus elements are modified as
\begin{align}
    \mathcal{L}'_U \eta &= \phi_{21} \left\{ [\psi_{11} U, \psi_{12} \eta]_A + \tilde{\mathcal{L}}_{\psi_{21} U} \psi_{12} \eta + \tilde{\mathcal{K}}_{\psi_{22} \eta} \psi_{11} U + R(\psi_{21} U, \psi_{22} \eta) \right\} \nonumber\\
    & \quad \ + \phi_{22} \left\{ [\psi_{21} U, \psi_{22} \eta]_Z + \mathcal{L}_{\psi_{11} U} \psi_{22} \eta + \mathcal{K}_{\psi_{12} \eta} \psi_{21} U + H(\psi_{11} U, \psi_{12} \eta) \right\} \, , \nonumber\\
    \mathcal{K}'_V \omega &= \phi_{21} \left\{ [\psi_{12} \omega, \psi_{11} V]_A + \tilde{\mathcal{L}}_{\psi_{22} \omega} \psi_{11} V + \tilde{\mathcal{K}}_{\psi_{21} V} \psi_{12} \omega + R(\psi_{22} \omega, \psi_{21} V) \right\} \nonumber\\
    & \quad \ + \phi_{22} \left\{ [\psi_{22} \omega, \psi_{21} V]_Z + \mathcal{L}_{\psi_{12} \omega} \psi_{21} V + \mathcal{K}_{\psi_{11} V} \psi_{22} \omega + H(\psi_{12} \omega, \psi_{11} V) \right\} \, ,
\end{align}
whereas the tilde-calculus elements are modified according to
\begin{align}
    \tilde{\mathcal{L}}'_\omega V &= \phi_{11} \left\{ [\psi_{12} \omega, \psi_{11} V]_A + \tilde{\mathcal{L}}_{\psi_{22} \omega} \psi_{11} V + \tilde{\mathcal{K}}_{\psi_{21} V} \psi_{12} \omega + R(\psi_{22} \omega, \psi_{21} V) \right\} \nonumber\\
    & \quad \ + \phi_{12} \left\{ [\psi_{22} \omega, \psi_{21} V]_Z + \mathcal{L}_{\psi_{12} \omega} \psi_{21} V + \mathcal{K}_{\psi_{11} V} \psi_{22} \omega + H(\psi_{12} \omega, \psi_{11} V) \right\} \, , \nonumber\\
    \tilde{\mathcal{K}}'_\eta U &= \phi_{11} \left\{ [\psi_{11} U, \psi_{12} \eta]_A + \tilde{\mathcal{L}}_{\psi_{21} U} \psi_{12} \eta + \tilde{\mathcal{K}}_{\psi_{22} \eta} \psi_{11} U + R(\psi_{21} U, \psi_{22} \eta) \right\} \nonumber\\
    & \quad \ + \phi_{12} \left\{ [\psi_{21} U, \psi_{22} \eta]_Z + \mathcal{L}_{\psi_{11} U} \psi_{22} \eta + \mathcal{K}_{\psi_{12} \eta} \psi_{21} U + H(\psi_{11} U, \psi_{12} \eta) \right\} \, .
\end{align}
Choosing tilde-calculus, $Z$-bracket and $R$-twist as 0 in the initial bracket, we can reduce these results of the general approach to the one yielded by Equation (\ref{halmagyitrick}) with replacing $\Psi_{\Pi}$ with an arbitrary $\Psi$ of the form (\ref{psiform}). 

At this point, we focus on three important types of automorphisms as $\Psi$ itself is rather general and not very instructive to discuss. In principle, one can perform an axiomatic analysis using the expressions displayed above, and the results that we give in Section \ref{s5} for a general transformation~$\Psi$. The first type that we consider is an extension of a frame change, the second one is an extension of $B$-transformation and the third one is a generalization of the $\beta$-transformation \cite{hohm2013spacetime} which are given in the form 
\begin{equation}
\Psi_m = \begin{pmatrix}
m  & 0 \\
0 & \tilde{m}
\end{pmatrix} \, , \qquad    
\Psi_B = \begin{pmatrix}
1 & 0 \\
B & 1
\end{pmatrix} \, , \qquad    
\Psi_\Pi = \begin{pmatrix}
1  & \Pi \\
0 & 1
\end{pmatrix} \, ,
\end{equation}
with the corresponding inverses:
\begin{equation}
\Psi_m^{-1} = \begin{pmatrix}
m^{-1}  & 0 \\
0 & \tilde{m}^{-1}
\end{pmatrix} \, , \qquad 
\Psi_B^{-1} = \begin{pmatrix}
1 & 0 \\
- B & 1
\end{pmatrix}\, , \qquad   
\Psi_\Pi^{-1} = \begin{pmatrix}
1  & - \Pi \\
0 & 1
\end{pmatrix} \, ,
\end{equation}
respectively. For the first case with $\Psi_m$, the new anchor reads 
\begin{equation}
    \rho_A' \oplus \rho_Z' = \rho_A m \oplus \rho_Z \tilde{m} \, ,
\end{equation}
and the brackets on $A$ and $Z$ are modified as
\begin{align}
    [U, V]'_A &= m^{-1} [m U, m V]_A \, , \nonumber\\
    [\omega, \eta]'_Z &= \tilde{m}^{-1} [\tilde{m} \omega, \tilde{m} \eta]_Z \, ,
\end{align}
where these imply that $m: (A, [\cdot,\cdot]_A) \to (A, [\cdot,\cdot]'_A)$ and $\tilde{m}: (Z, [\cdot,\cdot]_Z) \to (Z, [\cdot,\cdot]'_Z)$ are bracket morphisms. The modifications of $H$- and $R$-twists read
\begin{align}
    H'(U, V) &= \tilde{m}^{-1} H(m U, m V) \, , \nonumber\\
    R'(\omega, \eta) &= m^{-1} R(\tilde{m} \omega, \tilde{m} \eta) \, .
\end{align}
Moreover, we get the modification of the calculus elements as
\begin{align}
    \mathcal{L}'_U \eta &= \tilde{m}^{-1} \mathcal{L}_{m U} \tilde{m} \eta \, , \nonumber\\
    \mathcal{K}'_V \omega &= \tilde{m}^{-1} \mathcal{K}_{m V} \tilde{m} \omega \, , \nonumber\\
    \tilde{\mathcal{L}}'_{\omega} V &= m^{-1} \tilde{\mathcal{L}}_{\tilde{m} \omega} m V \, , \nonumber\\
    \tilde{\mathcal{K}}'_{\eta} U &= m^{-1} \tilde{\mathcal{K}}_{\tilde{m} \eta} m U \, .
\end{align}
Notice that, using $\Psi_m$ one cannot generate new terms to the original bracket. Particularly, if we choose the twisted Dorfman bracket as our initial bracket, we have $\tilde{\mathcal{L}} = \tilde{\mathcal{K}} = [\cdot,\cdot]_Z = R = 0$, and we cannot induce any term of the twisted tilde-Dorfman bracket via this $\Psi_m$ automorphism.

For the second case $\Psi_B$, the new anchor becomes 
\begin{equation}
    \rho_A' \oplus \rho_Z' = \rho_A + \rho_Z B \oplus \rho_Z \, ,
\end{equation}
and the brackets of $A$ and $Z$ are modified as
\begin{align}
    [U, V]'_A &= [U, V]_A + \tilde{\mathcal{L}}_{B U} V + \tilde{\mathcal{K}}_{B V} U + R(B U, B V) \, , \nonumber\\
    [\omega, \eta]'_Z &= [\omega, \eta]_Z - B R(\omega, \eta) \, .
\end{align}
The modifications of $H$-twist reads
\begin{align}
    H'(U, V) &= H(U, V) + \mathcal{L}_U B V  + \mathcal{K}_V B U + [B U, B V]_Z \nonumber\\
    & \quad - B [U, V]_A - B \tilde{\mathcal{L}}_{B U} V - B \tilde{\mathcal{K}}_{B V} U - B R(B U, B V) \, ,
\end{align}
where we observe that the $R$-twist does not get modified. Moreover, we get the modification of the calculus elements as 
\begin{align}
    \mathcal{L}'_U \eta &= \mathcal{L}_U \eta + [B U, \eta]_Z - B \tilde{\mathcal{K}}_{\eta} U - B R(B U, \eta) \, , \nonumber\\
    \mathcal{K}'_V \omega &= \mathcal{K}_V \omega + [\omega, B V]_Z - B \tilde{\mathcal{L}}_{\omega} V - B R(\omega, B V) \, , \nonumber\\
    \tilde{\mathcal{L}}'_{\omega} V &= \tilde{\mathcal{L}}_{\omega} V + R(\omega, B V) \, , \nonumber\\
    \tilde{\mathcal{K}}'_{\eta} U &= \tilde{\mathcal{K}}_{\eta} U + R(B U, \eta) \, .
\end{align}
If we want to start with only the twisted Dorfman bracket as our initial bracket, we choose $\tilde{\mathcal{L}} = \tilde{\mathcal{K}} = [\cdot,\cdot]_Z = R = 0$, then we see that the calculus elements $\mathcal{L}, \mathcal{K}$ and the bracket on $A$ do not get modified. Moreover, one cannot induce any term of the twisted tilde-Dorfman bracket via $\Psi_B$ when the initial bracket is the twisted Dorfman bracket. Yet the $H$-twist is modified by $- B[U, V]_A + \mathcal{L}_U B V + \mathcal{K}_V B U$, which reduces to $d B$ when the calculus is the usual Cartan calculus. This is a remnant of the \v{S}evera classification theorem for exact Courant algebroids. The map $B$ corresponds to a splitting for the exact sequence (\ref{exactcourant}), and it is a fundamental object in generalized geometry \cite{hitchin2010lectures}. Moreover it appears in low energy effective description of string theories as Kalb-Ramond 2-form field with field strength $H = dB$ which is related to the $H$ flux. 

Finally for $\Psi_\Pi$, the new anchor becomes 
\begin{equation}
    \rho_A' \oplus \rho_Z' = \rho_A \oplus \rho_Z + \rho_A \Pi \, ,
\end{equation}
and the brackets on $A$ and $Z$ are modified as
\begin{align}
    [U, V]'_A &= [U, V]_A - \Pi H(U, V) \, , \nonumber\\
    [\omega, \eta]'_Z &= [\omega, \eta]_Z + \mathcal{L}_{\Pi \omega} \eta + \mathcal{K}_{\Pi \eta} \omega + H(\Pi \omega, \Pi \eta) \, .
\end{align}
The modification of the $R$-twist reads
\begin{align}
    R'(\omega, \eta) &= R(\omega, \eta) + [\Pi \omega, \Pi \eta]_A + \tilde{\mathcal{L}}_{\omega} \Pi \eta + \tilde{\mathcal{K}}_{\eta} \Pi \omega \nonumber\\
    & \quad \ - \Pi [\omega, \eta]_Z - \Pi \mathcal{L}_{\Pi \omega} \eta - \Pi \mathcal{K}_{\Pi \eta} \omega - \Pi H(\Pi \omega, \Pi \eta) \nonumber\\
    &= R(\omega, \eta) + [\Pi \omega, \Pi \eta]_A + \tilde{\mathcal{L}}_{\omega} \Pi \eta + \tilde{\mathcal{K}}_{\eta} \Pi \omega - \Pi [\omega, \eta]'_Z \, ,
\end{align}
where we see that the $H$-twist does not get modified. Lastly, we get the modifications of the calculus elements as
\begin{align}
    \mathcal{L}'_U \eta &= \mathcal{L}_U \eta + H(U, \Pi \eta) \, , \nonumber\\
    \mathcal{K}'_V \omega &= \mathcal{K}_V \omega + H(\Pi \omega, V) \, , \nonumber\\
    \tilde{\mathcal{L}}'_\omega V &= \tilde{\mathcal{L}}_\omega V + [\Pi \omega, V]_A - \Pi \mathcal{K}_V \omega - \Pi H(\Pi \omega, V) \, , \nonumber\\ 
    \tilde{\mathcal{K}}'_\eta U &= \tilde{\mathcal{K}}_\eta U + [U, \Pi \eta]_A - \Pi \mathcal{L}_U \eta - \Pi H(U, \Pi \eta) \, .
\end{align}
If we want to start with only the twisted Dorfman bracket as our initial bracket, we choose $\tilde{\mathcal{L}} = \tilde{\mathcal{K}} = [\cdot,\cdot]_Z = R = 0$, then we see that the calculus elements $\mathcal{L}, \mathcal{K}$ and the bracket on $A$ are only modified with an $H$-twist term. Moreover, we see that one can still induce every term of the twisted tilde-Dorfman bracket. Particularly, one has
\begin{align}
    \tilde{\mathcal{L}}'_\omega V &= [\Pi \omega, V]_A - \Pi (\mathcal{K}_V \omega + H(\Pi \omega, V)) \, , \nonumber\\ 
    \tilde{\mathcal{K}}'_\eta U &= [U, \Pi \eta]_A - \Pi( \mathcal{L}_U \eta + H(U, \Pi \eta)) \, , \nonumber\\
    [\omega, \eta]'_Z &= \mathcal{L}_{\Pi \omega} \eta + \mathcal{K}_{\Pi \eta} \omega + H(\Pi \omega, \Pi \eta) \, , \nonumber\\
    R'(\omega, \eta) &= [\Pi \omega, \Pi \eta]_A - \Pi [\omega, \eta]'_Z \nonumber\\ 
    &= [\Pi \omega, \Pi \eta]_A - \Pi \left( \mathcal{L}_{\Pi \omega} \eta + \mathcal{K}_{\Pi \eta} \omega + H(\Pi \omega, \Pi \eta) \right) \, .
    \label{zazaoba}
\end{align}
We note that in general, on the Drinfel'd double $E = TM \oplus \Lambda^p T^*M$ there are contributions up to cubic order in $\Pi$ as in \cite{halmagyi2009non} due to the $R'$-twist, and the full bracket coincides with the higher Roytenberg bracket \cite{jurvco2013p, drinfeldpaper}. We can directly check the $C^{\infty} M$-linearity properties of the $R'$-twist. It is $C^{\infty} M$-linear in the second entry as it should be, but in the first entry we have
\begin{equation}
    R'(f \omega, \eta) = f R'(\omega, \eta) + L_A(f, \omega, \eta) - \Pi (\sigma_d(f, \iota_{\Pi \omega} \eta + \iota_{\Pi \eta} \omega)) \, . 
\end{equation}
Hence given $L_A = 0$ as in the Lie algebroid case, we deduce that if $R'$-twist is $C^{\infty}M$-linear in the first entry then $\Pi$ should have some non-trivial kernel. 

When we choose $A = TM, Z = \Lambda^p T^*M$ and $\Pi$ as a Nambu-Poisson structure, $\tilde{\mathcal{L}}'$ and $\tilde{\mathcal{K}}'$ in Equation (\ref{zazaoba}) coincide with the calculus elements we construct in \cite{drinfeldpaper} in the absence of the $H$-twist. Moreover the bracket on $\Lambda^p T^*M$ reduces to the Koszul bracket (\ref{koszulbracket}) \cite{koszul1985crochet, bi2011higher}, and the $R'$-twist vanishes due to the Nambu-Poisson condition. These calculus elements reduce to Poisson calculus \cite{vaisman2012lectures} for the case~$p = 1$. A ``dual'' geometric framework of generalized geometry using Poisson structures is introduced as Poisson generalized geometry in \cite{muraki2015new, asakawa2015poisson} by using the tilde-Dorfman bracket analogously to the usual Dorfman bracket (see also \cite{blumenhagen2013non}) depending on the exact sequence
\begin{equation} 
    0 \rightarrow TM \rightarrow E \rightarrow T^*M \rightarrow 0 \, .
\end{equation} 
In Poisson generalized geometry, one starts with the cotangent Lie algebroid equipped with the Koszul bracket defined in terms of a Poisson bivector. This bracket induces the Poisson calculus as a tilde-calculus which is dual to the usual Cartan calculus in the sense of Section \ref{s4} \cite{drinfeldpaper}. In turn, one can use the Koszul bracket and tilde-calculus to define a tilde-Dorfman bracket. Starting from this bracket as the initial bracket, one can use $\Psi_\Pi$ defined in terms of a new bivector $\Pi$ as a twist automorphism. In this dual setting, the bivector $\Pi$ takes a role analogous to Kalb-Ramond $B$-field in generalized geometry. Its field strength related to the $R$ flux \cite{asakawa2015poisson} corresponds to $\hat{d} \Pi$ where $\hat{d} \, \cdot := [\Pi,\cdot]_{\text{SN,Lie}}$ is the differential operator of the initial Poisson calculus. The differential $\hat{d}$ coincides with our $\tilde{d}$ when acting on smooth functions. In \cite{drinfeldpaper}, we extend this geometric setup to Nambu-Poisson exceptional generalized geometry and proved some preliminary results for arbitrary~$p$-forms. Since Poisson generalized geometry provides a useful framework for several crucial concepts from physics; including the non-geometric $R$ \cite{asakawa2015poisson} and $Q$ fluxes \cite{asakawa2015topological}, D-branes and Seiberg-Witten maps \cite{asakawa2012d}, we expect Nambu-Poisson exceptional generalized geometry to be fruitful for brane physics \cite{bagger2007modeling, bagger2008gauge}. This framework is also closely related to exceptional Drinfel'd algebras \cite{sakatani2020u, malek2020poisson}, which are U-duality extension of Drinfel'd doubles of Lie bialgebras.

We finally would like to note an important property of twist matrices; they take values in the symmetry group associated with the compactification scheme. Particularly for DFT this is given by $O(d,d)$, and for ExFT, $E_{d(d)}$. They are compatible with the fiber metric on the algebroid realizing the symmetries. In fact these considerations restrict the form of the twist matrices. Generally, twist matrices are locally defined; when they can be globally defined this is an example of a generalized parallelizable background \cite{grana2009t, du2018generalized}. As twist matrices are local, and take values in a group, they may be associated with a cohomological data defining the global structure of an algebroid. This is studied for DFT in \cite{coimbra2011supergravity} and for ExFT in \cite{coimbra2014e_d}. Moreover a similar idea is used for studying global structure of Leibniz algebroids associated with closed form symmetries in \cite{baraglia2012leibniz}. Also in \cite{asakawa2015poisson}, $\beta$-transformations are used for obtaining a global picture for Poisson generalized geometry defined in terms of a tilde-Dorfman bracket. In our earlier work \cite{drinfeldpaper}, in addition to Nambu-Poisson exceptional generalized geometry, we also studied global structures through an extension of formal rackoids of \cite{ikeda2021global} for vector bundle valued metrics. In a future work, we plan to investigate global aspects of algebroids in relation with exceptional Drinfel'd algebras as well.

%%%%%%%%%%%%%%%%%%%%%%%%%%%%%%%%%%%%%%

\section{Examples from the Literature}
\label{s8}

In Section \ref{s6}, we generalized the notion of proto Lie bialgebroids \cite{roytenberg2002quasi} to proto bialgebroids in concordance with our previous work \cite{drinfeldpaper} about the calculus framework on arbitrary vector bundles of the form $A \oplus Z$, where both $H$- and $R$-twists are present. In this section, we give several examples of these constructions. These include higher Courant \cite{bi2011higher}, conformal Courant \cite{baraglia2013conformal}, $AV$-Courant \cite{libland2011}, exact $E$-Courant \cite{Chen_2010}, Atiyah \cite{baraglia2013transitive} algebroids and $B_n$-generalized geometry \cite{baraglia2012leibniz}. The first three are rather simple and analogous, so we summarize them in the first subsection, whereas the others deserve a more detailed look in separate subsections. We are interested only in the brackets on these algebroids, and for the details we refer to each individual paper. We will not explicitly state which elements are in which vector bundle; we hope that it will be clear from the context. There are other crucial algebraic structures which can be incorporated in our calculus framework, namely exceptional Courant brackets \cite{pacheco2008m}, and exceptional Drinfel'd algebras \cite{sakatani2020u, malek2020poisson}. The simplest example of the latter inherits a Leibniz algebra structure on $TM \oplus \Lambda^2 T^*M$ in terms of a Nambu-Poisson trivector. The structure constants and compatibility conditions are directly related to the tilde-calculus elements constructed in the previous section for $\Psi_{\Pi}$ as observed in \cite{drinfeldpaper}. This construction yields a matched pair of Leibniz algebras and constitute an example of a bialgebroid since there are no twists. The exceptional Courant bracket on $TM \oplus \Lambda^2 T^*M \oplus \Lambda^5 T^*M$ is given by
\begin{equation}
    \left[U + \omega_2 + \omega_5, V + \eta_2 + \eta_5 \right]_E = [U, V]_{\text{Lie}} \oplus \left( \mathcal{L}_U \eta_2 - \iota_V d \omega_2 \right) \oplus \left( \mathcal{L}_U \eta_5 - \iota_V d \omega_5 - \eta_2 \wedge d \omega_2 \right) \, ,
\end{equation}
in terms of usual Cartan calculus elements. For the decomposition $A = TM \oplus \Lambda^2 T^* M, Z = \Lambda^5 T^*M$, this bracket includes an $H$-twist given by
\begin{equation}
    H(U + \omega_2, V + \eta_2) = - \eta_2 \wedge d \omega_2 \, ,
\end{equation}
so that we have a quasi bialgebroid structure. Moreover for the decomposition $A = TM \oplus \Lambda^5 T^*M, Z = \Lambda^2 T^*M$, we get an $R$-twist given by
\begin{equation}
    R(\omega_2, \eta_2) = 0 \oplus - \eta_2 \wedge d \omega_2 \, ,
\end{equation}
so that we have a tilde-quasi bialgebroid structure. The remaining decomposition $A = TM, Z = \Lambda^2 T^*M \oplus \Lambda^5 T^*M$ does not produce any twists as already observed in \cite{drinfeldpaper}, so we have a bialgebroid structure for this case. The topics on exceptional structures including exceptional Courant brackets and exceptional Drinfel'd algebras deserve more attention, and we plan to focus on them in separate future papers. Now we move on to the mentioned examples from the literature.

%%%%%%%%%%%%%%%%%%%%%%%%%%%%%%%%%%%

\subsection{Basic Examples}
\label{s8a}

One canonically has the usual Dorfman bracket on $E = TM \oplus T^*M$ as we have discussed in the previous section. The usual Dorfman bracket can be twisted by a 1-form valued 2-form $H$: 
\begin{equation}
    [U + \omega, V + \eta]_{\text{Dor}}^H = [U, V]_{\text{Lie}} \oplus \mathcal{L}_U \eta - \mathcal{L}_V \omega + d \iota_V \omega + H(U, V) \, , 
\end{equation}
given in terms of the usual Cartan calculus elements. The metric invariance of this bracket for the metric 
\begin{equation}
    g_E(U + \omega, V + \eta) = \iota_U \eta + \iota_V \omega \, ,
\end{equation}
implies that $H$ is indeed a 3-form which can be seen from Equation (\ref{twistedmetricinv}) since $g_A = g_H = \tilde{\iota} = 0$. Moreover, the Jacobi identity forces $H$ to be closed, \textit{i.e.}, $d H = 0$, which can be inferred from Equation (\ref{twistjacobih}). $H$ is linear in both entries, and the calculus elements are just the usual Cartan calculus so that all twisted linearity conditions are satisfied. All other twisted compatibility conditions hold, since the bracket and metric on $Z$ vanish as well as tilde-calculus elements and $R$-twist. These observations imply that the triplet $(\mathcal{L}, \iota, d)$ defines a calculus in the sense we introduced. An $H$-twist can be obtained from an automorphism procedure of the form $\Psi_B$ for a 2-form $B$. In this case we have $H = d B$ so that its closedness is guaranteed by its exactness. All of these can be easily extended to the \textit{higher} Dorfman bracket on $E = TM \oplus \Lambda^p T^*M$ which has the same form as above. In this case, the $H$-twist is given by a $p$-form valued 2-form, where $(p+2)$-forms, which are the ones usually considered in the literature \cite{bi2011higher}, constitute a special case. For the higher case, $B$-field is a $(p+1)$-form, and a similar automorphism procedure can be done \textit{mutatis mutandis}.

The same type of brackets with different calculus elements also appear in various algebroids. For example any exact conformal Courant algebroid \cite{baraglia2013conformal}, which by definition come equipped with an $E$-connection $\nabla$ on $L$, is constructed on the bundle $E = TM \oplus (T^*M \otimes L)$ where $L$ is a line bundle, with the bracket
\begin{equation}
    [U + \omega, V + \eta]_{\text{Dor}}^H = [U, V]_{\text{Lie}} \oplus \mathcal{L}^{\nabla}_U \eta - \mathcal{L}^{\nabla}_V \omega + d^{\nabla} \iota_V \omega + H(U, V) \, . 
\end{equation}
The calculus elements are the Cartan calculus on $L$-valued forms, with $d^{\nabla}$ being the exterior covariant derivative defined by
\begin{align}
    (d^{\nabla} \omega)(V_1, \ldots, V_{k+1}) &:= \sum_{1 \leq i \leq k+1} (-1)^{i+1} \nabla_{V_i}(\omega(V_1, \ldots, \check{V_i}, \ldots, V_{k+1})) \nonumber\\
    & \quad \ + \sum_{1 \leq i < j \leq {k+1}} (-1)^{i + j} \omega([V_i, V_j]_{\text{Lie}}, V_1, \ldots, \check{V_i}, \ldots \check{V_j}, \ldots, V_{k+1}) \, .
\label{ej15}
\end{align}
Moreover, the Lie derivative $\mathcal{L}^{\nabla}$ can be defined by the Cartan magic formula with $d^{\nabla}$ and the interior product $\iota$. In this case, the pairing is $L$-valued so that we have $\mathbb{E} = L$ and isotropic splittings exist. Moreover, the $H$-twist is given by a $d^{\nabla}$-closed $L$-valued 3-form $H$, which is a $(T^*M \otimes L)$-valued 2-form before imposing metric invariance and Jacobi identity.

Similarly, $AV$-Courant algebroids \cite{libland2011} have the same type of brackets on $E = A \oplus (A^* \otimes V)$ for a Lie algebroid $A$ and a vector bundle $V$. The calculus elements are the Cartan calculus on $V$-valued forms on $A$. The pairing takes values in $V$, \textit{i.e.}, $\mathbb{E} = V$, and isotropic splittings exist. The $H$-twist is given by a closed $V$-valued 3-form $H$ on $A$, which is a $(A^* \otimes V)$-valued 2-form before imposing metric invariance and Jacobi identity.

All of these examples are special cases of quasi metric-Bourbaki bialgebroids where the relevant structures on $Z$ vanishes, and the doubled bracket is of the form of a twisted Dorfman bracket.

%%%%%%%%%%%%%%%%%%%%%%%%%%%%%%%%%

\subsection{Exact $\mathbb{E}$-Courant Algebroids}
\label{s8b}

Exact Courant algebroids of the form $TM \oplus T^*M$ can be extended to omni-Lie algebroids \cite{chen2010omni} which generalize omni-Lie algebras \cite{weinstein2000omni}. This can be achieved by considering the tangent bundle as the derivations of smooth functions, and then replacing functions with an arbitrary vector bundle $\mathbb{E}$. Then an omni-Lie algebroid is of the form $E = \mathfrak{D} \mathbb{E} \oplus \mathfrak{J} \mathbb{E}$, where $\mathfrak{D} \mathbb{E}$ and $\mathfrak{J} \mathbb{E}$ denote the covariant operator bundle and the first jet bundle of $\mathbb{E}$, respectively. These two vector bundles have an $\mathbb{E}$-valued non-degenerate pairing $\langle \cdot,\cdot \rangle_{\mathbb{E}}: \mathfrak{J} \mathbb{E} \otimes \mathfrak{D} \mathbb{E} \to \mathbb{E}$, and in this sense they are $\mathbb{E}$-dual \cite{chen2010omni}. Both of these vector bundles are Lie algebroids \cite{kosmann2002differential, crainic2005secondary}, and due to $\mathbb{E}$-duality, one can generate analogous Cartan calculus elements, which yield a calculus in the sense we have introduced. The omni-Lie algebroid $E = \mathfrak{D} \mathbb{E} \oplus \mathfrak{J} \mathbb{E}$ is then an algebroid equipped with the Dorfman bracket constructed from these calculus elements. As we prove in \cite{drinfeldpaper}, these algebroids can be considered as a metric-Bourbaki bialgebroid where the algebroid structures on $\mathfrak{J} \mathbb{E}$ are chosen to be vanishing. 

One can axiomatize the properties of omni-Lie algebroids to define $\mathbb{E}$-Courant algebroids \cite{Chen_2010} (in \cite{Chen_2010}, these are defined as $E$-Courant algebroids; we have to replace $E$ with $\mathbb{E}$ in order to have a consistent notation in this paper). These algebroids satisfy analogous $\mathbb{E}$-valued properties similar to the usual Courant algebroids. One can introduce $\mathbb{E}$-Lie bialgebroids \cite{Chen_2010}, and as we prove in \cite{drinfeldpaper}, these are also examples of metric-Bourbaki bialgebroids. 

In \cite{Chen_2010}, \v{S}evera classification theorem for exact Courant algebroids is extended for exact $\mathbb{E}$-Courant algebroids, where now the one-to-one correspondence is between isomorphism classes of exact $\mathbb{E}$-Courant algebroids and equivalence classes of admissible pairs on the omni-Lie algebroid $E = \mathfrak{D} \mathbb{E} \oplus \mathfrak{J} \mathbb{E}$. An admissible pair is defined as a pair of maps $(\varpi, \Theta)$, where $\varpi: \mathfrak{D} \mathbb{E} \times \mathfrak{D} \mathbb{E} \to \mathbb{E}$ ($\varpi$ is denoted by $\omega$ in \cite{Chen_2010}) and $\Theta: \mathfrak{D} \mathbb{E} \times \mathfrak{D} \mathbb{E} \to \mathfrak{J} \mathbb{E}$ such that $\Theta$ is a 2-cocyle of the Leibniz cohomology of $\mathfrak{D} \mathbb{E}$ with coefficients in $\mathfrak{J} \mathbb{E}$ satisfying
\begin{align}
    \Theta(U, U) &= U(\varpi(U, U)) \, , \nonumber\\
    \tfrac{1}{2} U(\varpi(V, V)) &= \langle \Theta(U, V), V \rangle_{\mathbb{E}} + \varpi([U, V]_{\mathfrak{D} \mathbb{E}}, V) \, ,
\label{admissiblepair}
\end{align}
for all $U, V \in \mathfrak{D} \mathbb{E}$, where $[\cdot,\cdot]_{\mathfrak{D} \mathbb{E}}$ is the Lie algebroid bracket on $\mathfrak{D} \mathbb{E}$, which is given by the usual commutator of two differential operators. Here, analogously to the vector fields, we denote the action of a derivation $U \in \mathfrak{D} \mathbb{E}$ on an element $\xi \in \mathbb{E}$ by $U(\xi)$. In this setting, the classification theorem of \cite{Chen_2010} implies that any exact $\mathbb{E}$-Courant algebroid is isomorphic to $E = \mathfrak{D} \mathbb{E} \oplus \mathfrak{J} \mathbb{E}$ with the bracket and the metric
\begin{align}
    [U + \omega, V + \eta]_E &= [U + \omega, V + \eta]_{\text{Dor}} + \Theta(U, V) \, , \nonumber\\
    g_E(U + \omega, V + \eta) &= \tfrac{1}{2} \left( \langle \omega, V \rangle_{\mathbb{E}} + \langle \eta, U \rangle_{\mathbb{E}} \right) + \varpi(U, V) \, .
\end{align}
Here, the Dorfman bracket is written in terms of the calculus elements constructed from the Lie algebroid structure on $\mathfrak{D} \mathbb{E}$. 

After choosing the vector bundles $A = \mathfrak{D} \mathbb{E}, Z = \mathfrak{J} \mathbb{E}, \mathcal{Z} = \mathbb{E}$, we may directly identify $H = \Theta$ and $g_H = \varpi$ by the first Equation in (\ref{admissiblepair}), so that $\mathscr{Z}=\mathbb{E}$ as well. Moreover, with the identifications $d_H = d_{\mathfrak{D} \mathbb{E}}, \pounds = \text{id}_{\mathfrak{D} \mathbb{E}}$, and $\iota_V \omega = \langle \omega, V \rangle_{\mathbb{E}}$, we see that any exact $\mathbb{E}$-Courant algebroid is a quasi metric-Bourbaki bialgebroid in the sense that we have introduced. More concretely, since both $\mathfrak{D} \mathbb{E}$ and $\mathfrak{J} \mathbb{E}$ are Lie algebroids we have $g_A = g_Z = 0$ so that $\mathbb{A}, \mathbb{Z}, \mathbb{D}_A$ and $\mathbb{D}_Z$ are irrelevant. Moreover, the tilde-calculus, the bracket on $Z$ and the $R$-twist vanish, hence $\mathcal{A}$, $\mathscr{A}$ and $d_R$ are irrelevant. We first note that the twisted linearity conditions and their duals are satisfied due to Cartan magic formula for calculus elements and Equations (\ref{admissiblepair}). We also observe that all calculus twisted compatibility conditions and Jacobi twisted compatibility conditions together with their duals are trivially satisfied. Our Equation (\ref{twistjacobih}), whose dual is trivially satisfied, is equivalent to the fact that $\Theta$ is a 2-cocyle in the Leibniz cohomology. The condition (\ref{twistedmetricinv}) is then equivalent to the second condition in the definition of admissible pairs (\ref{admissiblepair}). All other metric invariance compatibility conditions together with their duals are trivially satisfied with the choice $\tilde{\pounds}=0$, except the dual of Equation (\ref{metricinvcond2twist}) follows from the fact that $\iota$ squares to zero.

%%%%%%%%%%%%%%%%%%%%%%%%%%%%%%%%%%%%

\subsection{Atiyah Algebroids}
\label{s8c}

\noindent Atiyah algebroids are transitive Lie algebroids related to principal $G$-bundles which are defined by Atiyah in the study of existence of complex analytic connections \cite{atiyah1957complex}. They naturally appear in different areas of physics. For instance in theoretical mechanics, for a system with configuration space admitting symmetries as a principal $G$-bundle, they yield different Lie algebroid realizations of Lagrange and Hamilton mechanics, see \cite{grabowski2011geometric} and references therein. Moreover they are used in formulations of gauge theories on transitive Lie algebroids \cite{fournel2013formulation}. More relevant for the purposes of this paper, they are used in string theory applications. Particularly, they appear in the constructions of heterotic Courant algebroids proposed in \cite{baraglia2013transitive}. The main aim for this construction is to address the T-duality in heterotic string theory where one needs a gauge bundle with a connection for its formulation. In this case, the relevant vector bundle becomes a transitive Courant algebroid $E\oplus T^*M$ where $E\to M$ denotes an Atiyah Lie algebroid. We now summarize the basics of Atiyah Lie algebroids following \cite{baraglia2013transitive}. 

Let $\pi: P \to M$ be a principal $G$-bundle with a compact, connected, semi-simple Lie group $G$. Since $G$ has a transitive, free action on sections of $P$, it provides a well-defined quotient $E = TP/G$ whose sections correspond to $G$-invariant vector fields on $P$. This is a transitive Lie algebroid with the anchor $\rho_E: E \to TM$ induced by the differential of the projection $\pi_*: TP \to TM$ which is surjective, and the bracket which will be explained shortly. The kernel of the anchor is isomorphic to the adjoint bundle $\mathfrak{g}_P = P \times_{Ad_G} \mathfrak{g}$, defining an exact sequence of vector bundles over $M$:
\begin{equation}
    0 \longrightarrow \mathfrak{g}_P \longrightarrow E \xrightarrow{\ \rho_E \ } TM \longrightarrow 0 \, . 
\end{equation}
This exact sequence is termed as the Atiyah sequence associated to the principal $G$-bundle $P$ and $E$ is referred to as the Atiyah Lie algebroid \cite{mackenzie1987lie} after the inception of Lie algebroids \cite{pradines1967theorie}. A left splitting $\Gamma: E \to \mathfrak{g}_P$ of the above-displayed exact sequence corresponds to a principal connection in the sense of Ehresmann. The splitting gives a decomposition~$E = TM \oplus \mathfrak{g}_P$. The curvature of the principal connection $\Gamma$ is a $\mathfrak{g}_P$-valued 2-form on $P$ defined as $F = d \Gamma + \tfrac{1}{2}[\Gamma, \Gamma]_{\mathfrak{g}}$. The principal connection also induces a vector bundle connection $\nabla: TM \times \mathfrak{g}_P \to \mathfrak{g}_P$. In this setup, sections of the Atiyah algebroid is furnished with the bracket 
\begin{equation}
    [U + \omega, V + \eta]_E = [U, V]_{\text{Lie}} \oplus \nabla_U \eta - \nabla_V \omega - [\omega, \eta]_{\mathfrak{g}_P} - F(U, V) \, .   \label{atiyahbracket}
\end{equation}
Moreover, the Cartan-Killing metric on $\mathfrak{g}$ extends to a quadratic form on $E$ denoted by $g_E$ which is non-degenerate since $G$ is semi-simple, hence Atiyah Lie algebroids are examples of quadratic Lie algebroids.  

We now start analyzing our algebroid axioms by identifying $A=TM$ and $Z=\mathfrak{g}_P$. Looking at the bracket (\ref{atiyahbracket}) we see that the tilde-calculus and the $R$-twist vanish, so $\mathcal{A}$, $\mathscr{A}$ and $d_R$ are irrelevant. We have the brackets on $A = TM$ and $Z = \mathfrak{g}_P$ given by the Lie brackets $[\cdot,\cdot]_{\text{Lie}}$ and $- [\cdot,\cdot]_{\mathfrak{g}_P}$. The $H$-twist is given by $-F$, and the calculus elements read
\begin{equation}
    \mathcal{L}_U \omega = - \mathcal{K}_U \omega = \nabla_U \omega \, ,
\end{equation}
so that we have
\begin{equation}
    \iota_U \omega = 0 \, ,
\end{equation}
and $\mathcal{Z}$ and $d$ are irrelevant. 

We first observe that twisted linearity conditions and their duals for calculus elements hold, since $\iota$ vanishes, and $F$ is a 2-form. For the first calculus twisted compatibility condition in (\ref{calculusconditionstwisted}), the left-hand side is the curvature operator of the connection, whereas the right-hand side becomes $[F(U, V), \mu]_Z$ after deleting the vanishing terms. This is satisfied by construction due to the adjoint action on $Z=\mathfrak{g}_P$, where its dual holds trivially. The last two conditions in (\ref{calculusconditionstwisted}) are satisfied since $\iota = 0$, and their duals also trivially hold since tilde-calculus and $R$-twist vanish. The condition (\ref{jacobiUVWa}) about the Jacobiator of $A$ holds since the tilde-calculus vanishes, its dual also holds since the $R$-twist vanishes. Since $H$-twist is a $\mathfrak{g}_P$-valued 2-form, it is anti-symmetric, so $g_H = 0$, and $\mathscr{Z}$ and $d_H$ are irrelevant. Both $A$ and $Z$ are Lie algebroids, so $g_A = g_Z = 0$ and $\mathbb{D}_A, \mathbb{D}_Z$ and vector bundles $\mathbb{A}$ and $\mathbb{Z}$ are irrelevant. The condition (\ref{twistjacobih}) then follows from the Bianchi identity for the curvature 2-form, namely $d^{\nabla} F = 0$ where $d^{\nabla}$ is the exterior covariant derivative (\ref{ej15}) acting on Lie algebra valued forms, which explicitly reads
\begin{align}
    d^{\nabla} F (U, V, W) &= \nabla_U (F(V, W)) - \nabla_V (F(U, W)) + \nabla_W (F(U, V)) \nonumber\\ 
    &\quad - F([U, V]_{\text{Lie}}, W) + F([U, W]_{\text{Lie}}, V) - F([V, W]_{\text{Lie}}, U) \, . 
\end{align}
The first and second lines correspond to the left- and right-hand sides of Equation (\ref{twistjacobih}) with correct signs since $\mathcal{K} = - \mathcal{L} = - \nabla$. The dual of this condition trivially holds since $R = 0$. The first Jacobi twisted compatibility condition (\ref{comp1twist}) is equivalent to the fact that $\nabla$ is a derivation of the Lie bracket on $\mathfrak{g}_P$. This follows from the constancy of the structure functions and the Jacobi identity of the Lie bracket on $\mathfrak{g}_P$, where its dual is trivially satisfied. The other two (\ref{comp2twist}, \ref{comp3twist}) and their duals are also trivially satisfied. Metric invariance twisted compatibility conditions (\ref{twistedmetricinv} - \ref{metricinvcond2twist}) and their duals are all trivially satisfied because all the metric components are zero, so that relevant terms vanish and $\pounds$ and $\tilde{\pounds}$ are irrelevant. Therefore Atiyah algebroids are examples of quasi metric-Bourbaki bialgebroids.

%%%%%%%%%%%%%%%%%%%%%%%%%%%%%%%%%%%%

\subsection{$B_n$-Generalized Geometry}
\label{s8d}

$B_n$-generalized geometry refers to the study of geometric structures on $TM \oplus C^{\infty}M \oplus T^*M$, where the $C^{\infty}M$ part often is denoted by only $1$. It can be seen as a motivating toy model proposed in \cite{baraglia2012leibniz} for understanding more complicated vector bundles such as the ones coming from exceptional field theories as in (\ref{physicsbundles}) \cite{rubio2014generalized}. The bracket on $E = TM \oplus C^{\infty}M \oplus T^*M$ is given by
\begin{equation}
    [U + f + \omega, V + g + \eta]_E = [U, V]_{\text{Lie}} \oplus U(g) - V(f) \oplus \mathcal{L}_U \eta - \mathcal{L}_V \omega + d \iota_V \omega + g d f \, ,
\end{equation}
in terms of usual Cartan calculus elements, and the metric reads
\begin{equation}
    g_E(U + f + \omega, V + g + \eta) = \iota_U \eta + \iota_V \omega + f g \, ,
\end{equation}
which takes values in $C^{\infty}M$ so that we have $\mathbb{E} = C^{\infty}M$. This is an $SO(n+1, n)$-invariant pairing which is of type $B_n$ in the Cartan classification. This is the reason for its naming. 

As now the vector bundle consists of three summands, we have a freedom to choose $A$ and~$Z$. Different decompositions correspond to different calculus elements and consequently different proto bialgebroid structures. All of these various decompositions do coincide on the Drinfel'd double. But as we will observe, different decompositions will be relevant for different components and the related compatibility conditions will change. 

First, we consider the decomposition $A = TM$ and $Z = C^{\infty}M \oplus T^*M$. In this case, the tilde-calculus and both $H$- and $R$-twists vanish so that we choose $g_H = g_R = 0$. Therefore $\mathcal{A}, \mathscr{Z},\mathscr{A},d_H$ and $d_R$ are irrelevant. The bracket on $A$ coincides with the usual Lie bracket so that $g_A = 0$ and $\mathbb{A}$, $\mathbb{D}_A$ are irrelevant. The calculus elements are given in terms of the usual Cartan calculus elements, with $\mathcal{Z} = C^{\infty} M$, and acting on possibly two different degrees:
\begin{align}
    \mathcal{L}_U (g + \eta) &= \mathcal{L}_U g + \mathcal{L}_U \eta = U(g) + \mathcal{L}_U \eta \, , \nonumber\\
    \iota_V (f + \omega) &= \iota_V f + \iota_V \omega = \iota_V \omega \, , \nonumber\\ 
    d f &= d f \, . 
\end{align}
Here, and in the following decompositions in this subsection, the operators on the left-most side of the displayed equations are the calculus elements in the sense that we introduced, whereas the other ones are the usual Cartan calculus elements. We hope that this will not cause any confusion to the reader. We have a bracket on $Z = C^{\infty}M \oplus T^*M$ defined by
\begin{equation}
    [f + \omega, g + \eta]_Z = g d f \, ,
\end{equation}
where $d$ is the usual exterior derivative. This bracket has a symmetric part which can be decomposed as in Equation (\ref{symmetricpart}) with 
\begin{equation}
    g_Z(f + \omega, g + \eta) = f g \, , \qquad \qquad \mathbb{D}_Z = d \, ,
\end{equation}
so that we have $\mathbb{Z} = C^{\infty}M$.

Firstly, twisted linearity conditions and their duals are all satisfied since Cartan magic formula holds and both twists vanish. Next, we observe that the calculus twisted compatibility conditions (\ref{calculusconditionstwisted}) all hold by the usual Cartan calculus relations, and $H = 0$. Since the bracket on $A$ is the usual Lie bracket, whose Jacobiator vanishes, and $H = 0$, the conditions (\ref{jacobiUVWa}) and (\ref{twistjacobih}) trivially hold. Duals of (\ref{calculusconditionstwisted}), (\ref{jacobiUVWa}) and (\ref{twistjacobih}) trivially hold since the tilde-calculus elements and the $R$-twist vanish. The first Jacobi twisted compatibility condition (\ref{comp1twist}) is satisfied due to the $C^{\infty}M$-linearity property of $\mathcal{L}$ in the second slot and the fact that $\mathcal{L} d = d \mathcal{L}$. The second one (\ref{comp2twist}) follows from $d^2 = 0$. The third one (\ref{comp3twist}) holds after cancellations of trivial terms due to Cartan magic formula. Their duals trivially hold. The first metric invariance twisted compatibility condition (\ref{twistedmetricinv}) holds trivially. The second condition (\ref{metricinvcond1twist}) holds by the identity for the commmutator of $\iota$ and $\mathcal{L}$ after setting $\pounds = \mathcal{L}$. Their duals holds with the identification $\tilde{\pounds} = 0$. The last condition (\ref{metricinvcond2twist}) follows from the Leibniz rule for the product of two functions. Its dual holds because $\iota$ squares to zero. Hence, we prove that for this decomposition we get a metric-Bourbaki bialgebroid structure.

Next, we consider the decomposition $A = TM \oplus C^{\infty}M$ and $Z = T^*M$. In this case, the tilde-calculus, the bracket on $Z$ and the $R$-twist vanish. Therefore we set $g_Z = 0$ and $\mathcal{A}, \mathbb{Z}, \mathscr{A}, \mathbb{D}_Z$ and $d_R$ are irrelevant. The bracket on $A$ is given by
\begin{equation}
    [U + f, V + g]_A = [U, V]_{\text{Lie}} \oplus U(g) - V(f) \, ,
\end{equation}
which is anti-symmetric so that $g_A = 0$, and the vector bundle $\mathbb{A}$ and the map $\mathbb{D}_A$ are irrelevant. The calculus elements are given in terms of the usual Cartan calculus element but acted by a vector and a function, with $\mathcal{Z} = C^{\infty}M$:
\begin{align}
    \mathcal{L}_{U + f} \eta &= \mathcal{L}_U \eta \, , \nonumber\\
    \iota_{U + f} \eta &= \iota_U \eta \, , \nonumber\\
    d \omega &= d \omega \, .
\end{align}
Moreover, the $H$-twist is given by
\begin{equation}
    H(U + f, V + g) = g d f \, ,
\end{equation}
whose symmetric part can be decomposed as in Equation (\ref{HRdecomp}) with
\begin{equation}
    g_H(U + f, V + g) = g f \, , \qquad \qquad \qquad d_H = d \, ,
\end{equation}
so that we have $\mathscr{Z} = C^\infty M$. 
 
With this setup, we observe that twisted linearity conditions and their duals hold due to Cartan magic formula and the fact that $H$-twist is linear in its second. All calculus twisted compatibility conditions (\ref{calculusconditionstwisted}) also hold because the left-hand sides vanish due to the usual Cartan calculus relations, whereas the right-hand sides vanish because the tilde-calculus and the bracket on $Z$ are absent. Their duals trivially hold since tilde-calculus and $R$-twist vanish. The condition for the Jacobiator of $A$ (\ref{jacobiUVWa}) holds by the commutator of the usual Lie derivative with itself. Its dual trivially holds. The other condition (\ref{twistjacobih}) holds by $d \mathcal{L} = \mathcal{L} d$, and its dual is trivially satisfied. All of the Jacobi twisted compatibility conditions (\ref{comp1twist} - \ref{comp3twist}), and their duals hold trivially as well since everything vanishes. The first of metric invariance twisted compatibility conditions (\ref{twistedmetricinv}) holds after identifying $\pounds_{U + f} = \mathcal{L}_U$ and using the action of Lie derivative on products of functions. Its dual is trivially satisfied. The second one (\ref{metricinvcond1twist}) holds with $\pounds_{U + f} = \mathcal{L}_U$ and using the identity for commutator of $\mathcal{L}$ and $\iota$, whose dual dictates $\tilde{\pounds} = 0$. The third one (\ref{metricinvcond2twist}) trivially holds, whereas its dual follows from the the fact that $\iota$ squares to zero after setting $\tilde{\pounds} = 0$. Hence, we prove that for this decomposition we get a quasi metric-Bourbaki bialgebroid structure. 

Lastly, we consider the decomposition $A = TM \oplus T^*M$, and $Z = C^{\infty}M$. In this case, the tilde-calculus elements, the bracket on $Z$ and the $H$-twist vanish. Therefore we set $g_H = g_Z = 0$ and $\mathcal{A}, \mathbb{Z}, \mathscr{Z}, \mathbb{D}_Z, d_H$ are irrelevant. The bracket on $A$ coincides with the usual Dorfman bracket:
\begin{equation}
    [U + \omega, V + \eta]_A = [U, V]_{\text{Lie}} \oplus \mathcal{L}_U \eta - \mathcal{L}_V \omega + d \iota_V \omega \, ,
\end{equation}
in terms of the usual Cartan calculus elements, whose symmetric part is given in terms of
\begin{equation}
    g_A(U + \omega, V + \eta) = \iota_U \eta + \iota_V \omega \, , \qquad \qquad \qquad \mathbb{D}_A = d \, ,
\end{equation}
with $\mathbb{A} = C^\infty M$. The only non-trivial calculus element is $\mathcal{L}$ which is given by
\begin{equation}
    \mathcal{L}_{U + \omega} f = - \mathcal{K}_{U + \omega}f = \mathcal{L}_U f = U(f) \, ,
\end{equation}
so that we have 
\begin{equation}
    \iota_{U + \omega} f = 0 \, ,
\end{equation}
and $\mathcal{Z}$ and $d$ are irrelevant. The $R$-twist is given by 
\begin{equation}
    R(f, g) = g d f \, ,
\end{equation}
whose symmetric part decomposes as in Equation (\ref{HRdecomp}) with 
\begin{equation}
    g_R(f, g) = f g \, , \qquad \qquad \qquad d_R = d \, ,
\end{equation}
so that we have $\mathscr{A} = C^\infty M$.

With this setup, all twisted linearity conditions and their duals hold because $\iota$ vanishes and $R$-twist is linear in its second slot. Moreover the calculus twisted compatibility conditions (\ref{calculusconditionstwisted}) are trivially satisfied by the usual Cartan calculus relations so that the left-hand sides vanish, and $H = 0$ so that the right-hand sides vanish. Their duals follow from Cartan magic formula and distributive property of Lie and exterior derivatives on products of smooth functions. The condition for the Jacobiator of $A$ (\ref{jacobiUVWa}) holds since the Dorfman bracket satisfies the Jacobi identity and $H = 0$, whereas its dual trivially holds. The condition (\ref{twistjacobih}) and its dual also hold trivially. All Jacobi twisted compatibility conditions (\ref{comp1twist} - \ref{comp3twist}) and their duals are trivially satisfied. The first metric invariance twisted compatibility condition (\ref{twistedmetricinv}) becomes the metric invariance property for the Dorfman bracket, and its dual trivially holds. The second condition (\ref{metricinvcond1twist}) holds trivially whereas for its dual, one needs to use $\mathbb{A} = \mathscr{A} = C^\infty M$ and 
\begin{equation}
    g_R(g, \mathcal{K}_{W + \mu} f) = - g_A(R(f, g), W + \mu) \, ,
\end{equation}
together with the identification $\tilde{\pounds} = 0$. The third condition (\ref{metricinvcond2twist}) of metric invariance twisted compatibility conditions holds after identifying $\pounds_{U + \omega} = \mathcal{L}_U$ and using the action of Lie derivative on products of functions. Its dual is equivalent to the metric invariance of the Dorfman bracket, which automatically holds. Hence, we prove that for this decomposition we get a tilde-quasi metric-Bourbaki bialgebroid structure.

In this subsection, we have discussed the algebroid structure on the triple sum vector bundle $E = TM \oplus C^{\infty}M \oplus T^*M$ which underlies the $B_n$-generalized geometry. As we observed, different decompositions of $E$ of the form $A \oplus Z$ lead to different calculus elements and relevant compatibility conditions change accordingly. They may even yield twistless or twistful cases as we explicitly evaluated. Yet, the consistency of the structures on the triple sum dictates that twisted compatibility conditions for every possible choice of such decompositions into two summands necessarily hold. This is particularly relevant for the higher dimensional exceptional geometries as we have seen in the beginning of this section. Moreover, heterotic Courant algebroids where one of the decompositions include the Atiyah algebroid of the previous subsection, are also of the form of a triple sum. A similar axiomatic approach of algebroid axioms on triple sum bundles coming from $O(d, d+n)$ gauged DFTs is studied in a recent work \cite{mori2024extended} based on their earlier results \cite{mori2020doubled, mori2020more}, which is also related to heterotic generalized geometry. They call the triple sum as an extended double and are interested in the case that two of the summands are Lie algebroids and a single one is a twisted Lie algebroid. Moreover they show that, equipped with a certain twisted C-bracket the extended double inherits a metric algebroid structure which is related to gauge algebra of DFT \cite{mori2020doubled}. Furthermore they show that the condition that the triple sum yields a Courant algebroid structure yields the physical section conditions of gauged DFT \cite{mori2024extended}. The conditions derived in \cite{mori2024extended} for various algebroid axioms in the level of Courant algebroids are directly relevant for our twisted compatibility conditions. It would be interesting to see the exact relationship between these two axiomatic approaches.

%%%%%%%%%%%%%%%%%%%%%%%%%%%%%%%%%%%%

\section{Concluding Remarks}
\label{s10}

This paper is a continuation of our earlier paper \cite{drinfeldpaper}, where we generalized Lie bialgebroids \cite{mackenzie1994lie} to bialgebroids. The former is defined on a pair of dual vector bundles whereas the latter is for a pair of arbitrary vector bundles. We achieved this generalization by introducing a framework of calculus on algebroids and studying certain algebroid properties of the Drinfel'd double structure on the direct sum of the pair. These properties include left- and right-Leibniz identities, Jacobi identity, symmetric part decomposition of the bracket, metric invariance, and properties that ensure that certain vector bundle maps are bracket morphisms. They dictate defining properties and impose compatibility conditions on two dual calculi. As there are many properties, there are many different algebroid structures. With this in mind, we loosely use the term bialgebroid freely without any adjectives. Our main result is that for a bialgebroid equipped with dual calculi satisfying a certain set of compatibility conditions, its Drinfel'd double satisfies the corresponding set of axioms. In particular we termed an algebroid which satisfies all properties as metric-Bourbaki algebroids. In this terminology, we showed that Drinfel'd double of a metric-Bourbaki bialgebroid is a metric-Bourbaki algebroid. 

In this paper, we extend the notion of bialgebroids to proto bialgebroids. Similarly to bialgebroids, we define proto bialgebroids on a pair of arbitrary vector bundles which are not necessarily dual and of arbitrary rank. In the former case, each algebroid forming the double is a subalgebroid. For proto bialgebroids this fact is relaxed to include the $H$- and $R$-twists. This allows us to study more general bracket structures on the Drinfel'd double, which is particularly useful for physical applications. Again, this generalization is obtained by an analysis of algebroid axioms similar to before. In general, the presence of twists mix different calculus elements and notions of two distinct calculi disappears in the sense of Section \ref{s4}. Nonetheless, the conditions imposed by each algebroid property are still interpreted as twisted compatibility conditions. In this case, the symmetric part decomposition of twists is new and is inspired by that of bracket. This decomposition already appears in examples in the literature and seems natural when one studies different decompositions associated with a triple sum. In particular, a symmetric part decomposition of bracket may become the symmetric part decomposition of twists when one uses a different decomposition. Examples to these fact are presented in Section \ref{s8} where we also present other examples of proto bialgebroids. Moreover, we defined proto metric-Bourbaki bialgebroids as proto bialgebroids whose Drinfel'd double satisfies all algebroid properties in the twistful case. 

By this study, we also achieve a second generalization, this time of proto Lie bialgebroids \cite{roytenberg2002quasi} which are generalizations of Lie bialgebroids to the twistful case. Similarly to Lie bialgebroids they are also defined on a pair of dual vector bundles. In Section \ref{s6} we proved that our definition of proto bialgebroids recovers that of proto Lie bialgebroids when appropriate choices are made. Let us make an important comparison at this point. For Lie bialgebroids, different definitions using the language of vector bundles and supermanifolds exist, and their equivalence is established. Similarly, proto Lie bialgebroids are defined in the language of supermanifolds \cite{roytenberg2002quasi} and first steps toward a bosonic definition, which is more suitable for physical applications, is carried out in \cite{chatzistavrakidis2015sigma}. Our analysis here is more detailed and general as we discuss in Section \ref{s6}. We present all our results in the language of vector bundles, and derive all conditions for arbitrary pairs of vector bundles which are not necessarily dual. This notion of proto bialgebroids extending proto Lie bialgebroids indeed coincide with the twistful extension of bialgebroids as indicated in our diagram. As we advocated throughout the paper, this is important for applications in ExFTs and U-dualities. With these motivations in mind, we also study the effects of vector bundle automorphisms on bracket structure of the Drinfel'd double. These automorphisms are intimately connected to twist matrices that appear in the physics literature and we comment on this relationship in Section \ref{s7}. 

In this paper, we analyzed two notions of twists; namely $H$- and $R$-twists and twist matrices in terms of an automorphism. There are other various twist notions that seem to be relevant to our work. One can twist the structures such as Poisson \cite{vsevera2001poisson}, Jacobi \cite{da2006twisted} or one can twist the calculi as in $d_H = d - H \wedge$ map \cite{witten1982supersymmetry}. More complicated fluxes may require more general twists \cite{shelton2005nongeometric, shelton2007generalized, blumenhagen2015relating, grana2009t}:
\begin{equation}
    d - H \wedge - f \circ - \ Q \bullet + R \ \llcorner \, .
\end{equation}
We plan to investigate such twists and their relation to our calculus formalism in the future. In our earlier study \cite{drinfeldpaper}, we constructed a calculus on $p$-forms dual to the usual Cartan calculus using Nambu-Poisson structures in the absence of twists. It would be interesting to look at the role of our calculus formalism on the twist notions of Nambu-Poisson structures. The naive generalization of twisted Nambu-Poisson structures by higher degree forms for multivectors does not work \cite{jurvco2013p}. Yet, we observe the instances of other types of twists. For example, for $E_{6(6)}$ Drinfel'd algebras \cite{malek2021e6}, the twist procedure is done by both a trivector and a hexavector. The 3-bracket induced by the trivector does not satisfy the fundamental identity; the identity is twisted by the hexavector. It is interesting to study such twists in a more general setting, and we plan to work on the details in the future. 

Exceptional Drinfel'd algebras in particular sustain an interesting territory for our calculus formalism. In the case of $SL(5)$ exceptional Drinfel'd algebras, we observe that the calculus elements constructed by the twist procedure with $\Psi_{\Pi}$ with Nambu-Poisson trivector $\Pi$ from the initial bracket chosen as the higher Dorfman bracket, which does not include $H$- or $R$-twists, reproduces the non-constant fluxes of \cite{sakatani2020u}. Moreover, the commutation relations are then induced by the Drinfel'd double structure and their quadratic constraints agree with our Jacobi compatibility conditions. {There are also generalizations of Drinfel'd algebras where the twists in the calculi are present \cite{rezaei2018jacobi}, which are relevant to extensions such as Jacobi-Lie T-plurality \cite{fernandez2021jacobi}. As we mentioned passing from T- to U-duality requires direct sum decompositions with not necessarily dual pairs. This generalization takes us upwards in our diagram. Exceptional Drinfel'd algebras are crucial for this transition as they are tailored for U-duality applications, and certain examples are constructed in \cite{sakatani2020u, malek2020poisson, malek2021e6, blair2022generalised, kumar202310}. On the other hand, there are various relevant studies which need a closer look. For example, Manin triples for U-duality are examined in \cite{musaev2021non}, and Nambu-Lie U-duality in the context of pullbacks are studied \cite{bugden2021g}. We believe that our formalism may be useful in providing a novel framework for a better understanding of exceptional Drinfel'd algebras, and pave the way for a better understanding of Nambu-Jacobi U-plurality.

As we have mentioned, there is also a closely relevant notion which is the exceptional Courant bracket \cite{pacheco2008m}. The complicated forms of higher dimensional examples may require one more extension for our calculus formalism. For example for $E_{7(7)}$, the relevant algebroid bracket is given by
\begin{align}
    \left[U + \omega_2 + \omega_5 + \omega_{1,7}, V + \eta_2 + \eta_5 + \eta_{1,7} \right]_E &= [U, V]_{\text{Lie}} \oplus \left( \mathcal{L}_U \eta_2 - \iota_V d \omega_2 \right) \oplus \left( \mathcal{L}_U \eta_5 - \iota_V d \omega_5 - \eta_2 \wedge d \omega_2 \right) \nonumber\\
    &\quad \oplus \left( \mathcal{L}_U \omega_{1,7} - j \eta_5 \wedge d \omega_2 - j \eta_2 \wedge d \omega_5 \right) \, ,
\label{obarey}
\end{align}
on the vector bundle
\begin{equation}
    E = TM \oplus \Lambda^2 T^*M \oplus \Lambda^5 T^*M \oplus \left( T^*M \otimes \Lambda^7 T^*M \right) \, .
\end{equation}
This bracket's symmetric part does not satisfy a decomposition property as in Equation (\ref{symmetricpart}), so they are examples of $Y$-algebroids \cite{hulik2024algebroids} which are not $G$-algebroids \cite{bugden2021g}. Extending the calculus formalism on such algebroids is of paramount importance for understanding exceptional geometries, and we plan to work on it in the future. These $Y$-algebroids are special cases of anti-commutable local Leibniz algebroids \cite{dereli2021anti}, on which we proved various fundamental geometrical identities, including Bianchi identities and Schouten decomposition, for certain ``admissible'' connections. 

Finally, we would like to have a better understanding of the connection between the supermanifold formalism \cite{voronov2012q, roytenberg2002quasi, roytenberg2007aksz, arvanitakis2018brane} and our calculus framework. Particularly, Cartan calculus operations on a manifold can be realized by certain supervectors on the cotangent bundle seen as a supermanifold. Since derived brackets \cite{Kosmann_Schwarzbach_2004} are used to construct generalizations of Cartan calculus elements, we want to investigate their relations to our calculus formalism. This is particularly interesting because the master equation is the key for AKSZ-type sigma models \cite{alexandrov1997geometry} and we hope to relate our calculus framework to the super language of AKSZ dictionary. Sigma model consistency, algebroid axioms and Bianchi identities are closely tied to each other \cite{ikeda2003chern, chatzistavrakidis2019fluxes}, and Bianchi identities for the Courant sigma model can also be recovered in terms of pre-Roytenberg algebras \cite{blumenhagen2012bianchi}. Our preliminary observations indicate that we can study pre-Roytenberg algebras in the calculus framework that we presented in this paper. By reformulating them in a frame independent manner, we see that it is possible to extend this result to the realm of higher Roytenberg algebras \cite{jurvco2013p, drinfeldpaper}. It would be interesting to study implications of this at the level of exceptional Drinfel'd algebroids, and in particular for the case $SL(5)$. We hope to come back to such questions in a future work.

%%%%%%%%%%%%%%%%%%%%%%%%%%%%%%%%%

\section*{Acknowledgements}

We are thankful to Oğul Esen, Serkan Sütlü and their research group for fruitful discussions. K.D. is funded by BAP Postdoctoral Research Fellowship (DOSAP) of İstanbul Technical University with Project Number TAB-2021-43202. This study was supported by Scientific and Technological Research Council of Turkey (TUBITAK) under the Grant Number 121F123. The authors thank TUBITAK for their supports.

%%%%%%%%%%%%%%%%%%%%%%%%%%%%%%%%%

\section*{Appendix: Calculus on Lie Algebroids}

In this appendix, we start with a summary of usual Cartan calculus of differential forms. We continue with Cartan calculus on arbitrary Lie algebroids. We discuss certain axioms and analyze almost-Lie, pre-Lie and Lie algebroids. We then relate them to our own calculus framework. Some of these results are used in the main proof of Section \ref{s6} which completes our diagram.

The tangent bundle $TM$ carries a Lie algebra structure due to the usual \textit{Lie bracket} $[\cdot,\cdot]_{\text{Lie}}$, which is an anti-symmetric $\mathbb{R}$-bilinear map satisfying the Jacobi identity. The Lie bracket also satisfies the (right- and left-)Leibniz rule, so that $(TM, \text{id}_{TM}, [\cdot,\cdot]_{\text{Lie}})$ becomes a Lie algebroid. The Lie bracket of vector fields can be extended to the \textit{Lie derivative} $\mathcal{L}: T M \times \Lambda^p T^*M \to \Lambda^p T^*M$ and the \textit{exterior derivative} $d: \Lambda^{p-1} T^*M \to \Lambda^p T^* M$. Together with the \textit{interior product} $\iota: T M \times \Lambda^p T^*M \to \Lambda^{p-1} T^*M$, they constitute the \textit{Cartan calculus}. In terms of the definitions in Section \ref{s4}, due to the properties listed below $(\mathcal{L}, \iota, d)$ is a \textit{calculus} on $\Lambda^p T^*M$ induced by $TM$ for any natural number $p$. These maps satisfy the following $C^{\infty}M$-linearity properties
\begin{align}
    \mathcal{L}_V (f \omega) &= f \mathcal{L}_V \omega + V(f) \omega \, , \nonumber\\
    \mathcal{L}_{f V} \omega &= f \mathcal{L}_V \omega + d f \wedge \iota_V \omega \, , \nonumber\\
    \iota_{f V} \omega &= \iota_V (f \omega) = f \iota_V \omega \, , \nonumber\\
    d(f \omega) &= f d \omega + d f \wedge \omega \, , 
\label{cartancalculuslinearity} 
\end{align}
for all $f \in C^{\infty} M, V \in T M, \omega \in \Lambda^p T^*M$. Moreover, they satisfy the following \textit{Cartan calculus relations}:
\begin{align}
    \mathcal{L}_U \mathcal{L}_V \omega - \mathcal{L}_V \mathcal{L}_U \omega &= \mathcal{L}_{[U, V]_{\text{Lie}}} \omega \, , 
    \nonumber\\
    \iota_U \iota_V \omega + \iota_V \iota_U \omega &= 0 \, ,
    \nonumber\\
    d^2 \omega &= 0 \, , 
    \nonumber\\
    \mathcal{L}_U \iota_V \omega - \iota_V \mathcal{L}_U \omega &= \iota_{[U, V]_{\text{Lie}}} \omega \, ,
    \nonumber \\
    \mathcal{L}_V d \omega - d \mathcal{L}_V \omega &= 0 \, ,
    \nonumber \\
    d \iota_V \omega + \iota_V d \omega &= \mathcal{L}_V \omega \, ,
\label{cartancalculusrelations} 
\end{align}
for all $U, V \in T M, \omega \in \Lambda^p T^*M$. 

Cartan calculus on the tangent Lie algebroid can be extended for an arbitrary almost-Lie algebroid $(A, \rho_A, [\cdot,\cdot]_A)$~\cite{mackenzie2005general}. The interior product $\iota: A \times \Lambda^p A^* \to \Lambda^{p-1} A^*$ is completely the same, whereas the Lie derivative $\mathcal{L}: A \times \Lambda^p A^* \to \Lambda^p A^*$ and exterior derivative $d: \Lambda^p A^* \to \Lambda^{p + 1} A^*$ can be defined by
\begin{align} 
    (\mathcal{L}_V \omega)(V_1, \ldots, V_p) &:= \rho_A(V)(\omega(V_1, \ldots, V_p)) - \sum_{i = 1}^p \omega(V_1, \ldots, [V, V_i]_A, \ldots V_p) \, ,
\label{liederivative} \\
    (d \omega)(V_1, \ldots, V_{p+1}) &:= \sum_{1 \leq i \leq p+1} (-1)^{i+1} \rho_A(V_i)(\omega(V_1, \ldots, \check{V_i}, \ldots, V_{p+1})) \nonumber\\
    & \quad \ + \sum_{1 \leq i < j \leq {p+1}} (-1)^{i + j} \omega([V_i, V_j]_A, V_1, \ldots, \check{V_i}, \ldots \check{V_j}, \ldots, V_{p+1}) \, ,
\label{exteriorderivative}
\end{align}
where $\check{V_i}$ indicates that $V_i$ is excluded.
These maps satisfy the same $C^{\infty}M$-linearity properties (\ref{cartancalculuslinearity}) with the only difference being in the first one where the second term on the right-hand side should read $\rho_A(V)(f) \omega$. When $A$ is a Lie algebroid, all Cartan relations (\ref{cartancalculusrelations}) are again satisfied. But relaxing the algebroid properties of $A$, we see that some of these conditions are not satisfied anymore. In particular, for a 1-form $\mu \in A^*$ we have the following properties
\begin{align}
    &\mathcal{L}_U \mathcal{L}_V \mu (W) - \mathcal{L}_V \mathcal{L}_U \mu (W) - \mathcal{L}_{[U, V]_A} \mu (W) = \mathcal{P}_{\rho_A}(U, V) (\iota_W \mu) - \mu (\mathcal{J}_A(U, V, W)) \, , \nonumber\\
    &d^2 \mu(U, V, W) = \mathcal{P}_{\rho_A}(U, V)(\iota_W \mu) - \mathcal{P}_{\rho_A}(U, W)(\iota_V \mu) + \mathcal{P}_{\rho_A}(V, W)(\iota_U \mu) - \mu(\mathcal{J}_A(U, V, W)) \, , \nonumber\\
    &\mathcal{L}_U d \mu (V, W) - d \mathcal{L}_U \mu (V, W) = \mathcal{P}_{\rho_A}(U, V)(\iota_W \mu) - \mathcal{P}_{\rho_A}(U, W)(\iota_V \mu) - \mu(\mathcal{J}_A(U, V, W)) \, ,
    \label{calculuslie-oneform}
\end{align}
whereas the others remain the same. Here, $\mathcal{J}$ denotes the Jacobiator and the \textit{predator} of the anchor of a vector bundle $E$ is defined by
\begin{equation}
    \mathcal{P}_{\rho_E}(u, v) := [\rho_E(u), \rho_E(v)]_{\text{Lie}} - \rho_E([u, v]_E) \, .
\end{equation}
Consequently right-hand sides of these three properties do not vanish for almost-Lie and pre-Lie algebroids, where they simplify for the latter. On the other hand, for a 0-form, \textit{i.e.}, a smooth function $f \in C^{\infty}M$, we have
\begin{align}
    \mathcal{L}_U \mathcal{L}_V f - \mathcal{L}_V \mathcal{L}_U f - \mathcal{L}_{[U, V]_A} f &= \mathcal{P}_{\rho_A}(U, V)(f) \, , \nonumber\\
        d^2 f(U, V) &= \mathcal{P}_{\rho_A}(U, V)(f) \, , \nonumber\\
        \mathcal{L}_U d f (V) - d \mathcal{L}_U f (V) &= \mathcal{P}_{\rho_A}(U, V)(f) \, ,
    \label{calculuslie-function}
\end{align}
and the others again remain the same. Hence, these three properties also remain the same for pre-Lie and Lie algebroids since the Jacobi identity together with the right-Leibniz rule implies that the anchor is a morphism of brackets (see for example Appendix B of \cite{drinfeldpaper} for the proof of this simple fact). Moreover, the calculus conditions from Section \ref{s4} read acting on 1-forms
\begin{align}
    \mathcal{L}_U d \iota_W \eta (V) - d\iota_{[U, W]_A} \eta (V) - d \iota_W \mathcal{L}_U \eta (V) &= \mathcal{P}_{\rho_A}(U, V) (\iota_W \eta) \, , \nonumber\\
    \mathcal{L}_W d \iota_V \omega (U) - d \iota_W d \iota_V \omega(U) &= \mathcal{P}_{\rho_A}(W, U) (\iota_V \omega) \, .
\label{calculuslie}
\end{align}
Hence, for pre-Lie and Lie algebroids right-hand sides vanish. Therefore the triplet satisfies the twistless versions of calculus conditions. These conditions also hold for a smooth function trivially for any almost-Lie algebroid since interior product acting on functions vanishes. The results that we presented above are used in the proof of Section \ref{s6} where we show proto Lie bialgebroids are examples of proto bialgebroids that we introduce here.

\newpage

\bibliographystyle{unsrt}
\bibliography{bibliography}

\end{document}